%% file: syskuramotovar.tex
\documentclass[review,onefignum,onetabnum]{siamart190516}

\input{syskuramotovar-shared}

\allowdisplaybreaks

\begin{document}
\newsiamthm{assumption}{Assumption}
\maketitle
\begin{abstract}
We look into the fluctuations caused by disturbances in power systems. In the linearized system of the power systems, the disturbance is modeled by a Brownian motion process, and the fluctuations are described by the covariance matrix of the associated stochastic process at the invariant probability distribution. We derive explicit formulas for the covariance matrix for the system with a uniform damping-inertia ratio. The variance of the frequency at the node with the disturbance is significantly bigger than the sum of those at all the other nodes, indicating the disturbance effects the node most, according to research on the variances in complete graphs and star graphs. Additionally, it is shown that adding new nodes typically does not aid in reducing the variations at the disturbance's source node. Finally, it is shown by the explicit formulas that the line capacity affect the variation of the frequency and the inertia affects the variance of the phase differences.
\end{abstract}
%
%
%
\section{Introduction}
\label{sec:intro}
A power system consists of synchronous machines, transmission lines and power supply and demand. 
The electricity system needs the frequency to be synchronized in order to operate properly. The frequencies of the synchronous machines (such as rotor-generators driven by steam or gas turbines) should all be equal to or near the nominal frequency (such as 50 Hz or 60 Hz) in a synchronous state of the power system \cite{Kundur1994}. Here, the frequency is the rotating phase angle's derivative, and it equals the synchronous machine's rotational speed, measured in rad/s. Synchronization stability, also known as transient stability in the field of power systems research, is defined as the capacity to retain synchronization under disturbances. The electrical system is experiencing an unprecedented threat of losing synchronization as a result of the expansion of the integration of renewable energy sources, which are inherently more vulnerable to unpredictable disturbances.
 \par 
Here, we focus on the relation of synchronuous stability with the variance of the disturbances. The
relation depends on the power system parameters in particular upon:
the inertia and the damping coefficients of the synchronous machines, the susceptance of the transmission lines, 
the power supply and demands and the network topology and so on. 
Based on the analysis of the existence condition \cite{DorflerCriticalcoupling,Jafarpour,skar_uniqueness_equilibrium}, the small signal stability \cite{motter} 
and the basin attraction of the synchronous state \cite{menck2,sizeofbasin,Xi2016}, the synchronization stability 
may be improved by changing these parameters, such as changing the inertia of the synchronous machines \cite{optimal_inertia_placement}, controlling the power flows in the network \cite{PIAC3}, adding or deleting transmission lines \cite{FAZLYAB2017181}. In the analysis, the focus is on the 
synchronous state itself, in which the disturbances have not yet been explicitly considered in the mathematical model. However, in practice, due to continuously occurring disturbances, the state always fluctuates around a synchronous state. If both 
the fluctuations in the frequency at the nodes and the phase angle differences between the nodes connected by lines are so large that the state cannot return to the basin attraction of the synchronous state, the synchronization is lost. 
Thus, the influences of the disturbances cannot be neglected and the severity of the fluctuations characterizes the synchronous stability. 
\par 
The $\mathcal{H}_2$ norm 
of an input-output linear system, in which the disturbances are modelled as input and the frequency deviation and 
the phase angle differences as output, has been used 
to measure the severity of the fluctuations \cite{RobustnessSynchrony,H2norm,optimal_inertia_placement}.  By minimizing this norm, parts of the system parameters 
can be assigned to suppress the fluctuations in the frequency and the phase angle differences. However, the $\mathcal{H}_2$ norm, which 
equals to the trace of a matrix,  is a global metric for the synchronization stability.  The fluctuations of the frequency at each node, the phase angle difference in each line and their correlation can hardly be explicitly characterized.  Clearly, the nodes with serious fluctuations in the frequencies and the lines with serious fluctuations in the phase angle differences are vulnerable to disturbances.  These nodes and lines cannot be effectively identified by the $\mathcal{H}_2$ norm. 
\par 
In physics,  the propagation of the fluctuations caused by 
the disturbances is investigated \cite{Propagation1,Delocalization,topological-spreading,Aue17,Zha19}. For example, 
the statistics of the fluctuations at the nodes, e.g., the variance of the increment of the frequency distribution, can be calculated via simulations by modelling the disturbances by either Gaussian or non-Gaussian noise \cite{Propagation1}. With perturbations added to the system parameters,  the disturbance arrival time and the vertex and edge susceptibility are estimated in \cite{topological-spreading,Networksuseptbilities} respectively.  The amplitude of perturbation responses of the states at the nodes are used to
study the emergent complex response patterns across the network \cite{Zha19}. 
By these investigations on fluctuations, intuitive insights on the impact of the system parameters, e.g., the network topology and the inertia of synchronous machines, on the spread of the disturbances are provided, which may help to develop practical guiding principles for real network design and control. 
\par 
In \cite{ZhenWangAutom},  the disturbance is modelled 
by a Brownian process in the linearized system of the nonlinear power systems and 
the fluctuations in the frequency and the phase angle differences are characterized by the 
variance matrix in the invariant probability distribution of the stochastic process. Formulas of the variance matrix 
have been deduced in \cite{ZhenWangAutom} with the assumption of uniform disturbance-damping among the nodes, in which 
the ratio of the strength of the disturbances and the damping coefficients are all identical at the nodes. By means of these formulas, the dependence of the fluctuations on the system parameters are investigated. 
Needed is an understanding of how the disturbances supplied to nodes propagate through the power
network and hence affect the phase angle differences and the frequencies of all nodes. Here, using this framework for studying the fluctuations in the system, we deduce the explicit formula for the variance matrix with an assumption of uniform damping-inertia ratios at the nodes and analyze the dependence 
of the propagation of the fluctuations from a node with a disturbance to the other nodes in the network. 
\par 
The contributions of this paper to the analysis of power systems include:
\begin{enumerate}[(i)]
\item with the assumption of the uniform damping-inertia ratios at the nodes,  we obtain the explicit formulas of the variance matrices of the frequency and the phase angle differences in lines;
\item based on the formulas, we analyse the dependence of the propagation of the disturbances on the system parameters in special graphs including 
complete graphs and star graphs. 
\end{enumerate} 
\par 
This paper is organized as follows. 
In Section \ref{Section:preliminary},  elementary preliminaries on graph theory 
and the invariant probability distribution of Gaussian process are provided. 
The problem formulation and the main results of this paper are presented in Section \ref{Section:problem} and \ref{main:results} respectively.
Section \ref{Section:proofs} provides proofs of the results and Section \ref{sec:conclusions} concludes with remarks. 

%
%
%
%
\section{Preliminaries}\label{Section:preliminary}
The elementary notation, properties of graphs and the concept of 
the asymptotic variance of a stochastic Gaussian system are introduced in this section. 
\subsection{Notations}
\par
The set of the integers is denoted by
$\mathbb{Z} = \{ \ldots, ~ -1, ~ 0, ~ 1, ~ 2, \ldots \}$
and that of the positive integers by
$\mathbb{Z}_+ = \{ 1, ~ 2, ~ \ldots \}$.
For any integer $n \in \mathbb{Z}$
denote the set of the first $n$ positive integers by
$\mathbb{Z}_n = \{ 1, ~ 2, ~ \ldots, ~ n \}$.
The set of the real numbers is denoted by $\mathbb{R}$.
Denote the strictly positive real numbers by $\mathbb{R}_{+} = (0, ~ +\infty)$.
\par
The vector space of $n$-tuples of the real numbers
is denoted by $\mathbb{R}^n$ for an integer $n \in \mathbb{Z}_+$.
For the integers $n, ~ m \in \mathbb{Z}_+$
the set of $n$ by $m$ matrices with entries of the real numbers,
is denoted by $\mathbb{R}^{n \times m}$.
Denote the identity matrix of size $n$ by $n$ by
$\mathbf{I}_n  \in \mathbb{R}^{n \times n}$, the zero vector by $\mathbf{0}_n$,
the vector with all elements equal to one by $\mathbf{1}_n$,
which may also be denoted by $\mathbf{I}$, $\mathbf{0}$ and $\mathbf{1}$ respectively
if the size is clear from the context. Denote the zero vector by $\mathbf{0}_n$ which may also 
be denoted by $\mathbf{0}$.
\par
Denote subsets of matrices according to:
for an integer $n \in \mathbb{Z}_+$,
$\mathbb{R}_{spd}^{n \times n}$ denotes
the subset of symmetric positive semi-definite matrices
of which an element is denoted by $0 \preceq \mathbf{Q} = \mathbf{Q}^{\top}$;
$\mathbb{R}_{ortg}^{n \times n}$
the subset of orthogonal matrices
which by definition satisfy 
$\mathbf{U} ~ \mathbf{U}^{\top} = \mathbf{I}_n = \mathbf{U}^{\top} ~ \mathbf{U}$.
Call a square matrix $\mathbf{A} \in \mathbb{R}^{n \times n}$ {\em Hurwitz}
if all eigenvalues have a real part which is strictly negative;
in terms of notation,
for any  eigenvalue 
$\lambda(\mathbf{A})$ of the matrix $\mathbf{A}$, 
$\realpart (\lambda(\mathbf{A})) < 0$. For a matrix $\mathbf{A}$, denote the element
at the entry $(i,j)$ by $a_{i,j}$. The common formula for the entries at position ($i,j$) of matrix $\mathbf{A}$  is denoted by $\mathbf{A}:a_{i,j}$. 

\par 
\subsection{Graphs}\label{subsection:graph}
Consider an undirected weighted graph $\mathcal{G}=(\mathcal{V},\mathcal{E})$
with a set of $n \in \mathbb{Z}_+$ nodes denoted by $\mathcal{V}$ and 
a set of $m \in \mathbb{Z}_+$ edges or lines denoted by $\mathcal{E}$ and line weight $w_{i,j}=w_{j,i}\in\mathbb{R}_{+}$ if the nodes $i$ and $j$ are connected  and $w_{i,j}=0$ otherwise. Denote by $k = (i, ~ j) \in \mathcal{E}$
the edge connecting the nodes $i$ and $j$
which edge is also denoted by $e_k$.
The Laplacian 
matrix  of the graph with weight $w_{i,j}$ of line $(i,j)$ is defined as  $\mathbf{L}=(l_{{i,j}})\in\mathbb{R}^{n\times n}$ with
 \begin{eqnarray}\label{LaplacianDef}         
          l_{i,j} =
          \left\{
          \begin{array}{ll}
            - w_{i,j},                               & \mbox{if} ~ i\neq j,\\
            \sum_{k=1, ~ k \neq i}^n ~ w_{i,k}& \mbox{if} ~ i=j.
          \end{array}  
          \right. \nonumber 
\end{eqnarray}
The incidence matrix is defined as $\widetilde{\mathbf{C}} =(c_{i,k}) \in \mathbb{R}^{n\times m}$ with $c_{i,k}\in\mathbb{R}$,
\begin{eqnarray}\label{IncidenceDef}  
        c_{i,k} 
    & = & \left\{
          \begin{array}{rl}
             1, & \text{if node $i$ is the beginning of line $e_k$},\\
            -1, & \text{if node $i$ is the end of line $e_k$},\\
             0, & \text{otherwise},
          \end{array}  
          \right.
          \end{eqnarray}
 Here the direction of line $e_k$ is arbitrarily specified in order to define the incidence matrix. 
 Elementary properties of matrices, which are needed subsequently, are summarized in 
 the next lemma.
 \begin{lemma}\label{lemma:laplacianmatrixproperties}
Consider the graph $\mathcal{G}$ and its Laplacian matrix $\mathbf{L}$.
\begin{itemize}
\item[(i)] The Laplacian matrix $\mathbf{L}$ is symmetric and hence all its eigenvalues are real.
\item [(ii)] Following the Gerschgorin' theorem  \cite[Theorem 36]{PietVanMieghem2008}, all the eigenvalues of $\mathbf{L}$ are non-negative. 
\item[(iii)] Denote the eigenvalues of $\mathbf{L}$ by $0\leq\mu_1\leq\mu_2\leq\cdots\leq \mu_n$. 
It holds
$\mathbf{L}  \mathbf{1}_n = \mathbf{0}_n$, thus, $\mu_1=0$ is an eigenvalue of $\mathbf{L}$ with an eigenvector $\tau \mathbf{1}_n$ where $\tau\in\mathbb{R}$.
\item [(iv)] The graph $\mathcal{G}$ is connected if and only if the second smallest eigenvalue $\mu_2>0$ \cite[Theorem 10]{PietVanMieghem2008}.  
 \end{itemize}
\end{lemma}
\par
\begin{figure}\label{Fig_Network}
\centering
\includegraphics[scale=0.4]{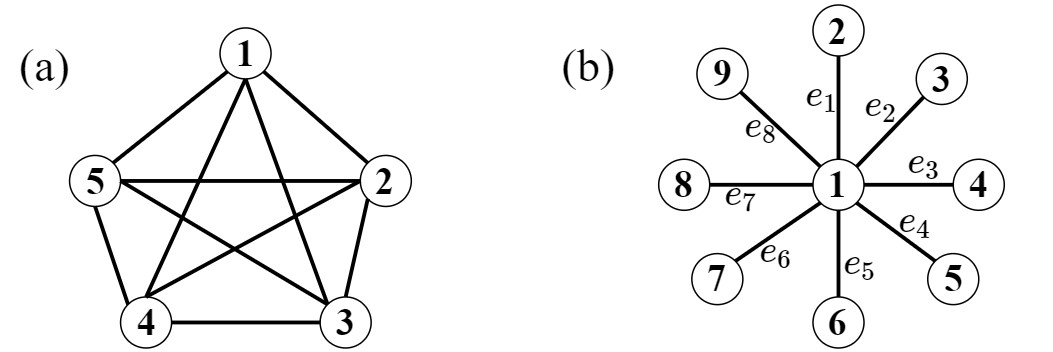}
\caption{(a) A complete graph with 5 nodes. (b) A star graph with 9 nodes.}
\end{figure}

 
The definitions of complete graphs and star graphs are described below.
\begin{definition}
Consider the graph $\mathcal{G}=(\mathcal{V},\mathcal{E})$. 
\begin{itemize}
\item[(i)] If each pair of nodes is connected
by a line, then call this graph a complete graph. 
\item [(ii)] If the graph is a tree and there is a root node which is directly connected
to all the other nodes, then  call this graph a star graph. 
\end{itemize}
\end{definition}
For both a complete graph and a star graph, the form of the incidence matrix depends on the indices of the lines.
For convenience of expression, we define the indices for the nodes 
and lines as below.
\begin{definition}\label{definition_index}
Consider the graph $\mathcal{G}=(\mathcal{V},\mathcal{E})$.
\begin{itemize}
\item[(i)]If $\mathcal{G}$ is a complete graph, then the indices 
of the line $(i,j)$ with $i<j$ is defined according to the Lexicographic order. 
\item[(ii)] If $\mathcal{G}$ is a star graph, the index of the 
root node is defined as $i=1$ and the indices of the other nodes are defined
as $i=2,\cdots,n$. The indices of the line $(1,k+1)$ are defined as $e_{k}$ for $k=2,\cdots,n-1$.
\end{itemize} 
\end{definition} 
An example of a complete graph and  an example of a star graph with such indices are shown in Fig.~\ref{Fig_Network}.
For the complete graph and the star graph, we have the following lemma.
\begin{lemma}\label{lemma_eigenvalues}
Consider the graph $\mathcal{G}=(\mathcal{V},\mathcal{E})$. Assume the weights of all the lines are all identical, 
i.e., $w_{i,j}=\nu\in\mathbb{R}_+$ for $(i,j)\in\mathcal{E}$, 
\begin{itemize}
\item[(i)] If $\mathcal{G}$ is a complete graph, then the eigenvalues of the Laplacian matrix satisfy \cite{PietVanMieghem2008}, 
\begin{align*}
\mu_1=0, ~\text{and}~\mu_i=\nu n ~\text{for}~ i=2,\cdots,n.
\end{align*}
 In addition, the incidence matrix has the following form,
\begin{equation*}
    \mathbf{\widetilde{C}}=
    \begin{bmatrix}
        1 & 1 & 1 & \cdots & 1 & 0 &\cdots & 0\\
        -1 & 0 & 0 & \cdots & 0 & 1 &\cdots & 0\\
        0 & -1 & 0 & \cdots & 0 & -1 &\cdots & 0\\
        0 & 0 & -1 & \cdots & 0 & 0 &\cdots & 0\\
        \vdots & \vdots &\vdots & \vdots &\vdots & \vdots & \vdots &\vdots \\
        0 & 0 & 0 & \cdots & -1& 0 & \cdots &-1
    \end{bmatrix}. 
\end{equation*}

\item[(ii)] If $\mathcal{G}$ is a star graph, then  the eigenvalues of the Laplacian matrix satisfy  \cite{PietVanMieghem2008}, 
\begin{align*}
\mu_1=0,\mu_2=\cdots=\mu_{n-1}=\nu, \mu_n= \nu n,
\end{align*}
the vector 
$\begin{bmatrix}
n-1&-1&-1&\cdots&-1
\end{bmatrix}^{\top}\in\mathbb{R}^n$ is an eigenvector of the Laplacian matrix
corresponding to the eigenvalue $\mu_n=\nu n$. In addition, 
with the indices defined in Definition \ref{definition_index}, the incidence 
matrix has the following form,
        \begin{equation*}
            \mathbf{\widetilde{C}}=
            \begin{bmatrix}
                1 & 1 & 1 & \cdots & 1\\
                -1 & 0 & 0 & \cdots & 0\\
                0 & -1 & 0 & \cdots & 0\\
                0 & 0 & -1 & \cdots & 0\\\
                \vdots & \vdots &\vdots & \vdots &\vdots\\
                0 & 0 & 0 & \cdots & -1
            \end{bmatrix}. 
        \end{equation*}
\end{itemize}
\end{lemma}
\par

\par 

\subsection{The Asymptotic Variance}\label{subsection:asymptotic}
Consider a time-invariant linear stochastic differential equation
with representation,
\begin{eqnarray*}
    \text{d} \mathbf{x}(t)
    & = & \mathbf{A} \mathbf{x}(t)  \text{d}t 
          + \mathbf{M} d \mathbf{v}(t), ~ \mathbf{x}(0) = \mathbf{x}_0, \\
        \mathbf{y}(t)
    & = & \mathbf{N} \mathbf{x}(t), 
\end{eqnarray*}
where $\mathbf{x}: \Omega \times T \rightarrow \mathbb{R}^{n_x}$; $\mathbf{A} \in \mathbb{R}^{n_x \times n_x}$;
 $\mathbf{M} \in \mathbb{R}^{n_x \times n_v}$; $\mathbf{v}: \Omega \times T \rightarrow \mathbb{R}^{n_v}$, is a standard Brownian motion with $\mathbf{v}(t) - \mathbf{v}(s) \in G(0, \mathbf{I}_{n_v} (t-s)), \forall ~ t, ~ s \in T, ~ s< t$; $\mathbf{x}_0 \in G(0, \mathbf{Q}_{\mathbf{x}_0})$ with $\mathbf{Q}_{\mathbf{x}_0}\in \mathbb{R}_{spd}^{n_x \times n_x}$
 is a Gaussian random variable; $\mathbf{y}: \Omega \times T \rightarrow \mathbb{R}^{n_y}$, $\mathbf{N} \in \mathbb{R}^{n_y \times n_x}$. 
  A standard Brownian motion is a stochastic process 
which starts at $t=0$ with $\mathbf{v}(0) = \mathbf{0}$,
has independent increments,
and the probability distribution of each increment is specified by
$(\mathbf{v}(t) - \mathbf{v}(s)) \in G(0, ~ (t-s)\mathbf{I}_{n_v})$ for any $s, ~ t \in T$ with $s < t$, meaning that 
$(\mathbf{v}(t)-\mathbf{v}(s))$ has a Gaussian probability distribution with mean zero and variance $(t-s)\mathbf{I}_{n_v}$. 

It follows from 
\cite[Theorem 1.52]{linearOptimalSystems} and
\cite[Theorem 6.17]{karatzas}
that the state process $\mathbf{x}$ and the output process $\mathbf{y}$
are Gaussian processes.
Denote then for all $t \in T$,
$\mathbf{x}(t) \in G(\mathbf{m}_x(t), ~ \mathbf{Q}_{x, tv}(t))$ with $\mathbf{Q}_{x, tv}(t)\in \mathbb{R}_{spd}^{n_x \times n_x}$ and 
$\mathbf{y}(t) \in G(\mathbf{m}_y(t), ~ \mathbf{Q}_{y, tv}(t))$ with $\mathbf{Q}_{y, tv}(t)\in \mathbb{R}_{spd}^{n_y \times n_y}$.
If in addition the matrix $\mathbf{A}$ is Hurwitz
then there exists an invariant probability distribution
of this linear stochastic system 
with the representation and properties
\begin{eqnarray*}
\mathbf{0}
    & = & \lim_{t \rightarrow \infty} ~ \mathbf{m}_x(t), ~
          \mathbf{0} = \lim_{t \rightarrow \infty} ~ \mathbf{m}_y(t), 
          \nonumber \\
        \mathbf{Q}_x 
    & = & \lim_{t \rightarrow \infty} ~ \mathbf{Q}_{x, tv}(t), 
          \mathbf{Q}_y = \lim_{t \rightarrow \infty} ~ \mathbf{Q}_{y, tv}(t), 
\end{eqnarray*}
where the variance matrix
\begin{eqnarray*}
\mathbf{Q}_x
    & = & \int_0^{+\infty} \exp(\mathbf{A} t) 
          \mathbf{M} \mathbf{M}^{\top}
          \exp(\mathbf{A}^{\top} t) 
          \text{d}t,~~\mathbf{Q}_y=\mathbf{N}\mathbf{Q}_x\mathbf{N}^{\top}.
\end{eqnarray*}
Here $\mathbf{Q}_x$ is the unique solution of the matrix equation
\begin{eqnarray}
 \mathbf{0}
    & = & \mathbf{A}  \mathbf{Q}_x 
          + \mathbf{Q}_x \mathbf{A}^{\top} 
          + \mathbf{M} \mathbf{M}^{\top}. 
          \label{eq:lyapunoveq}
\end{eqnarray}
One calls the matrix
$\mathbf{Q}_x$ the {\em asymptotic variance of the state process} and 
$\mathbf{Q}_y$ the {\em asymptotic variance of the output process} and the matrix equation (\ref{eq:lyapunoveq})
the {\em (continuous-time) Lyapunov equation}
for the asymptotic variance $\mathbf{Q}_x$.
Because the matrix $\mathbf{A}$ is assumed to be Hurwitz,
this equation has a unique solution
which can be computed by a standard iterative procedure.
In general the solution $\mathbf{Q}_x$ 
is symmetric and positive semi-definite.
If the matrix tuple $(\mathbf{A}, ~ \mathbf{M})$ is a controllable pair
then the matrix $\mathbf{Q}_x$ is positive definite,
denoted by $0 \prec \mathbf{Q}_x$.
These results may be found in
\cite[Theorem 1.53, Lemma 1.5]{linearOptimalSystems} and
\cite{karatzas}.

\section{Problem Formulation}\label{Section:problem}
In this section, we present the model of the power system and formulate the problem.  
\par 
The power network can be modelled by a graph $\mathcal{G}(\mathcal{V},\mathcal{E})$ with nodes
$\mathcal{V}$ and edges $\mathcal{E}\subset \mathcal{V}\times\mathcal{V}$, where a node represents a bus and an edge $(i,j)$ represents the transmission line between nodes $i$ and $j$. We focus on the transmission network and assume the lines are lossless. We denote the number of nodes in $\mathcal{V}$ and edges in $\mathcal{E}$ by $n$ and $m$, respectively. 
The dynamics of the power systems are described in the following definition.
\begin{definition}
Consider an undirected graph 
$\mathcal{G} = (\mathcal{V},\mathcal{E})$ 
with a set of $n \in \mathbb{Z}_+$ nodes denoted by $\mathcal{V}$ and 
a set of $m \in \mathbb{Z}_+$ edges or lines denoted by $\mathcal{E}$.
The system of the power system
is described by the dynamics \cite{zab2,menck2,hirsch1},  
\begin{subequations}\label{Nonlinear}
\begin{align}
\dot{\delta}_i&=\omega_i,\\
m_i\dot{\omega}_i&=P_i-d_i\omega_i-\sum_{j=1}^nK_{i,j}\sin{(\delta_i-\delta_j)},
\end{align}
\end{subequations}
where $\delta_i$ and $\omega_i$ denote the phase angle and the frequency deviation of the synchronous machine at node $i$; $m_i>0$ describes the inertia of the synchronous generators; $P_i$ denotes power generation 
if $P_i>0$ and denotes power load otherwise; 
$K_{i,j}=\hat{b}_{i j}V_iV_j$ is the effective susceptance, where $\hat{b}_{i,j}$ is the susceptance of the line $(i,j)$, $V_i$ is the voltage; $d_i>0$ is the damping coefficient with droop control. 
\end{definition}
\par 
In this definition, the dynamics of the voltage is not considered, which is assumed to be constant. This
is practical because the voltage can be controlled in a short time-scale thus can be approximated as constant 
in the time-scale of the frequency. 
\par 
When the graph is complete, and $d_i=1$ for all the nodes 
and $K_{i,j}=K/n$ for all $(i,j)\in\mathcal{E}$ 
with  $K\in\mathbb{R}_{+}$,
the system becomes \emph{the second-order Kuramoto Model}
\cite{Dorfler20141539}. 
\begin{definition}\label{Def_synchronous}
Define a {\em synchronous state}
of the power system (\ref{Nonlinear}) 
as the vector $\big(\delta^*(t),\omega^*(t)\big)$ with $\mathbf{\delta}^*(t)=\widetilde{\mathbf{\delta}}+(\widetilde{\omega} t)\mathbf{1}_n \in \mathbb{R}^n$ and $\omega^*(t)=\widetilde{\omega}\mathbf{1}_n\in\mathbb{R}^n$, 
which is a solution of the equation
\begin{eqnarray}\label{flows}
        d_i\widetilde{\omega}
    & = &  P_i +
          \sum_{j=1}^{n}~ K_{i,j} ~ \sin(\widetilde{\delta}_j - \widetilde{\delta}_i ),~\text{for}~i=1,\cdots,n
\end{eqnarray}
and $\widetilde{\delta}=\text{col}(\widetilde{\delta}_i)\in\mathbb{R}^n$  that satisfies $\widetilde{\delta}_i-\widetilde{\delta}_j=(\delta_i^*(t)-\delta_j^*(t))(\emph{mod}(2\pi))$ for all $(i,j)\in\mathcal{E}$. 
\end{definition}
\par
 By summing all the equations in (\ref{flows}), it yields that at the synchronous state
 \begin{align}\label{synchronizedfrequency}
 \widetilde{\omega}=\frac{\sum_i^n{P_i}}{\sum_i^n{d_i}}\in\mathbb{R}.
 \end{align}
 \par 
 The existence of a  synchronous state can typically be obtained 
by increasing the coupling strength $K_{i,j}$ for all the lines
 to sufficiently high values \cite{DorflerCriticalcoupling}.  
 \par 
The derivation of the linearized system of (\ref{Nonlinear})
is briefly summarized below with an assumption for the synchronous state.
\begin{assumption}\label{assumption:networkedsystem}
Consider the system (\ref{Nonlinear}), assume that 
(1) the graph $\mathcal{G}$ is connected, hence $m \geq n-1$ holds;
(2) there exists a synchronous 
state $\big(\mathbf{\delta}^*(t),\mathbf{0})$ such that the phase differences $|\widetilde{\delta}_i-\widetilde{\delta}_j|<\pi/2$ for all $(i,j)\in\mathcal{E}$.
\end{assumption}
\par 
The {\em linearized system}  of (\ref{Nonlinear}),
linearized around the considered synchronous state, 
is then derived 
\begin{align}\label{Linearization}
\left(\begin{array}{c} {\dot{\bm{\delta}}} \\ {\bm{ \dot{\omega}}}\end{array}\right)&=\left(\begin{array}{cc} \ 0 &\mathbf{I}_n\\
-\mathbf{M}^{-1}\mathbf{L} & -\mathbf{M}^{-1}\mathbf{D}\end{array}\right) \left(\begin{array}{c} {\bm{\delta}} \\{\bm{ \omega}}\end{array}\right)= \mathbf{J} \left(\begin{array}{c} {\bm{\delta}} \\{\bm{ \omega}}\end{array}\right),
\end{align}
where $\bm{\delta}=\text{col}{(\delta_i)}\in\mathbb{R}^n$, $\mathbf{I}_n\in\mathbb{R}^{n\times n}$ is the identity matrix, 
$\bm{\omega}=\text{col}{(\omega_i)}\in\mathbb{R}^n$, $\mathbf{M}=\text{diag}(m_i)\in\mathbb{R}^{n\times n}$,  $\mathbf{D}=\text{diag}{(d_i)}\in\mathbb{R}^{n\times n}$, and $\mathbf{L}\in\mathbb{R}^{n\times n}$ is the Laplacian matrix of the graph with weight
\begin{align*} 
w_{i,j}=K_{i,j}\cos{\delta_{ij}^*},~\text{for the line}~(i,j),
\end{align*}
generated by $(\bm{\delta}^*,\mathbf{0})$ with $\delta_{ij}^*=\delta_i^*-\delta_j^*$, 
$\mathbf{J}\in\mathbb{R}^{2n\times 2n}$ is also called \emph{the Jacobian matrix} of the power system 
at the synchronous state. Note that the state variables in (\ref{Linearization}) are the deviations of the phase angles and frequencies from the synchronous state $(\bm{\delta}^*,\mathbf{0})$. By the second Lyapunov method, the stability of $(\bm{\delta}^*,\mathbf{0})$ can be 
determined by the sign of the real part of the eigenvalues of $\mathbf{J}$. The analysis 
of the eigenvalue of matrix $\mathbf{J}$  of (\ref{Linearization}) is also called \emph{small-signal stability analysis}.
It has been proven that if $K_{i,j}\cos{\delta_{ij}^*}> 0$, then the system is stable  at the synchronous state $(\bm{\delta}^*,\bm 0)$ \cite{bronski,zab1}, which leads to the 
security condition 
\begin{align}\label{condition}
\mathbf{\Theta}=\big\{\bm\delta\in\mathbb{R}^n \big |~|\delta_{ij}|< \frac{\pi}{2},\forall (i,j)\in \mathcal{E})\big\}.
\end{align}

%
 \par
 Similarly, as in \cite{Kaihua_SIAM}, we model the disturbance by a Brownian motion process, which is then
 the input to a linear system, and study the stochastic system 
\begin{subequations}\label{stochasticsystem}
\begin{align}
\text{d}\bm\delta(t)&=\bm \omega(t)\text{d}t,\\
\text{d}\bm \omega(t)&=-\mathbf{M}^{-1}\big(\mathbf{L}\mathbf{\delta}(t)+\mathbf{D}\bm \omega(t)\big )\text{d}t+\mathbf{M}^{-1}\widetilde{\bm B}\text{d}\bm{\text{v}}(t)
\end{align}
\end{subequations}
with the state variable, system matrix and input matrix, 
\begin{align*}
 \bm x=\begin{bmatrix}
     \bm \delta\\
     \bm\omega
    \end{bmatrix},~~
 \mathbf{A}=\begin{bmatrix}
     \bm 0 & \mathbf{I}_n\\
    -\mathbf{M}^{-1}\mathbf{L}&-\mathbf{M}^{-1}\mathbf{D}
   \end{bmatrix},~~
 \mathbf{B} &=\begin{bmatrix}
    \bm 0\\
    \mathbf{M}^{-1}\widetilde{\mathbf{B}}
   \end{bmatrix},
\end{align*}
where $\widetilde{\mathbf{B}}=\text{diag}(b_i)\in\mathbb{R}^{n\times n}$ with $b_i>0$ being the strength of the disturbances of node $i$; $\bm{\text{v}}(t)=\text{col}(\text{v}_i(t))\in\mathbb{R}^n$ where $\text{v}_i(t)$ is a Brownian motion process that results in Gaussian distributed incremental disturbances at the nodes. 
The noise components $\text{v}_1, ~ \text{v}_2, ~ \ldots, ~ \text{v}_n$
are assumed to be independent.
Here, we refer to $K_{i,j}$ as the
\emph{line capacity} of line $e_k$, which is also called the coupling strength between the synchronous machines, and refer to $w_{i,j}=K_{i,j}\cos{\delta_{ij}^*}$ 
as the \emph{weight} of line $e_k$. It is obvious that the weights of the lines are determined by the line capacity and the power flows at the synchronous state which is solved from (\ref{flows}). 
Note that the weight depends on the line capacity in a non-linear way, i.e., increasing the 
line capacities of the lines, the phase differences $\delta_{ij}^*$ may decrease which further increases
the weights of the lines.

\par
In the model (\ref{stochasticsystem}), the disturbances denoted by $\text{v}_i(t)$ at node $i$ are assumed to be independent, which is reasonable because the locations of the power generators, including renewable power generators, are usually far from each other. Because the system (\ref{stochasticsystem}) is linear, at any time, the probability distribution of the state is Gaussian. We focus on the variance matrices of the frequency and of the phase
angle difference in the invariant probability distribution of the
linear stochastic system, which reflect the dependence of the
fluctuations of the frequency and the phase angle difference on
the system parameters.  To focus on the fluctuations in the frequency 
and the phase angle differences,  when considering the variance matrix in the invariant probability distribution,
we set the output matrix so that 
\begin{align}\label{output-delta-omega}
\mathbf{y}=\mathbf{C}\mathbf{x},~~
\mathbf{y}=
\begin{bmatrix}
\mathbf{y}_\delta\\
\mathbf{y}_\omega
\end{bmatrix},
~~
\mathbf{C}=
\begin{bmatrix}
\widetilde{\mathbf{C}}^{\bm\top}&\bm 0\\
\bm 0&\mathbf{I}_n
\end{bmatrix}
\in\mathbb{R}^{(m+n)\times 2n}.
\end{align}
The $m$ elements in $\mathbf{y}_\delta$ are the phase angle differences
in the $m$ lines, and the $n$ elements in $\mathbf{y}_\omega$ are the frequencies 
at the $n$ nodes. The matrix $\widetilde{\mathbf{C}}=(c_{i,k})\in\mathbb{R}^{n\times m}$ is the incidence matrix of the graph $\mathcal{G}$. 
\par 
To study the dependence of the fluctuations in the frequency and 
the phase differences of the system (\ref{Nonlinear}) on the system parameters,  the asymptotic 
variance of the frequency and the phase difference in the system (\ref{stochasticsystem}) are 
investigated.  
Here, we denote the variance matrix of the output by 
\begin{align}\label{Q_y_form}
\mathbf{Q}_y=
\begin{bmatrix}
\mathbf{Q}_\delta& \mathbf{Q}_{\delta\omega}^{\bm\top}\\
\mathbf{Q}_{\delta\omega}&\mathbf{Q}_\omega
\end{bmatrix}\in\mathbb{R}^{(m+n)\times(m+n)}, \mathbf{Q}_\delta\in\mathbb{R}^m, \mathbf{Q}_{\delta\omega}
\in\mathbb{R}^{n\times m}, \mathbf{Q}_\omega\in\mathbb{R}^{n\times n}. 
\end{align}
For comparison with the main result of this paper,  we present 
the asymptotic variance of the state in the Single-Machine Infinite Bus (SMIB) model, which is governed by the dynamics,
\begin{subequations}
\begin{align}
\dot{\delta}&=\omega,\\
\eta\dot{\omega}&=P-d\omega-K\sin{\delta},
\end{align}
\end{subequations}
Assume there exists a synchronous state $(\arcsin{(P/K)},0)$.  
The linear stochastic system of SMIB model corresponding to the system (\ref{stochasticsystem}) is 
\begin{subequations}\label{Linearization11}
\begin{align}
\text{d}\delta(t)&= \omega(t)\text{d}t,\\
\text{d} \omega(t)&=-\eta^{-1}\big(l\delta(t)+d\omega(t)\big )\text{d}t+\eta^{-1}\beta\text{d}\bm{\text{v}}(t)
\end{align}
\end{subequations}
where $l=K\cos{\delta^*}=\sqrt{K^2-P^2}$. We set the output as $y=(\delta,\omega)^{\bm \top}$.
By solving a Lyapunov function,
 \begin{align*}
  &\mathbf{A}\mathbf{Q}_x+\mathbf{Q}_x\mathbf{A}^{\bm\top}+\mathbf{B}\mathbf{B}^{\bm\top}=\bm 0, 
 \end{align*}
with 
\begin{align}
\mathbf{A}=
\begin{bmatrix}
0 &1\\
- \eta^{-1} l &- \eta^{-1}d
\end{bmatrix},~~
\mathbf{B}=
\begin{bmatrix}
0\\
 \eta^{-1}\beta
\end{bmatrix},
\end{align}
we obtain the variance matrix $\mathbf{Q}_y$ of the output 
\begin{align}\label{SMIBQ}
\mathbf{Q}_y=\mathbf{Q}_x=
\begin{bmatrix}
 \frac{\beta^2}{2d\sqrt{K^2-P^2}} &0\\
0 &\frac{\beta^2}{2\eta d}
\end{bmatrix}. 
\end{align}
From the explicit formula of $\mathbf{Q}_y$, it is found 
that the variance of the phase angle is independent on the inertia and 
the variance of the frequency is independent on the line capacity. 
The roles of the damping played on the suppression of the variance of 
the phase angle and the frequency are the same. 
Obviously, due to the simplicity of this model, the fluctuations
in the power networks with multi-machines cannot be fully explored by this model.

The problem of the characterization of the asymptotic variance of the stochastic linear system 
(\ref{stochasticsystem}) is described below.
\begin{problem}\label{problem}
Consider the stochastic linearized power system (\ref{stochasticsystem}) with multi-machines.
Deduce an analytic expression of the asymptotic variance
of the output process $\mathbf{y}$ and display how this variance depends on the system parameters.
\end{problem}
\par 
The theorem for the solution of Problem \ref{problem}
makes use of the properties and the notations in the following lemma.

\begin{lemma}\label{appendix_theorem}
Consider the Laplacian matrix $\mathbf{L}$ and the positive-definite diagonal matrix $\mathbf{M}$ in system (\ref{stochasticsystem}). There exists an orthogonal matrix $\mathbf{U}\in\mathbb{R}^{n\times n}_{ortg}$ such that
\begin{align}\label{decomposition}
\mathbf{U}^{\bm\top}\mathbf{M}^{-1/2}\mathbf{L}\mathbf{M}^{-1/2}\mathbf{U}=\mathbf{\Lambda}_n, 
\end{align}
where $\bm\Lambda_n=\text{diag}(\lambda_i)\in\mathbb{R}^{n\times n}$ with $0= \lambda_1< \lambda_2\cdots<\lambda_n$ being the eigenvalues of the matrix $\mathbf{M}^{-1/2}\mathbf{L}\mathbf{M}^{-1/2}$, $\mathbf{U}=
\begin{bmatrix}
\mathbf{u}_1&\mathbf{u}_2&\cdots&\mathbf{u}_n
\end{bmatrix}$ with $\bm u_i\in\mathbb{R}^n$ being the eigenvector corresponding to $\lambda_i$ for $i=1,\cdots,n$. In addition, $\bm u_1=1/\sqrt{n}\bm 1_n$.  
\end{lemma}
\par 
For the asymptotic variance matrix of the stochastic system (\ref{stochasticsystem}), we have the following theorem \cite{ZhenWangAutom}. 
\begin{theorem}\label{theoremmain0}
Consider the stochastic system (\ref{stochasticsystem}) with Assumption \ref{assumption:networkedsystem} and
the notations of matrices in Lemma  \ref{appendix_theorem}. 
Define matrices
\begin{equation}
\begin{aligned}
 &\mathbf{A}_e=\begin{bmatrix}
     \mathbf{0} & \mathbf{I}_n\\
    -\mathbf{\Lambda}_n &-\mathbf{U}^{\bm\top}\mathbf{M}^{-1}\mathbf{D}\mathbf{U}
   \end{bmatrix}\in\mathbb{R}^{2n\times 2n},  
\mathbf{B}_e =\begin{bmatrix}
    \mathbf{0}\\
    \mathbf{U}^{\top}\mathbf{M}^{-\frac{1}{2}}\widetilde{\mathbf{B}}
   \end{bmatrix}\in \mathbb{R}^{2n\times n},\\
   &\mathbf{C}_e=
\begin{bmatrix}
\widetilde{\mathbf{C}}^{\bm\top} \mathbf{M}^{-\frac{1}{2}}\mathbf{U}&\mathbf{0}\\
\mathbf{0}&\mathbf{M}^{-\frac{1}{2}}\mathbf{U}
\end{bmatrix}\in\mathbb{R}^{(m+n)\times 2n}, 
    \label{blockmatrix0}
\end{aligned}
\end{equation}
which which can be decomposed according to 
\begin{align}\label{blockmatrix}
 \mathbf{A}_e=\begin{bmatrix}
         \mathbf{0} & \mathbf{A}_{12}\\
     \mathbf{0} &\mathbf{A}_2
   \end{bmatrix}, ~~
  \mathbf{B}_e =\begin{bmatrix}
    \mathbf{0}\\
    \mathbf{B}_2
   \end{bmatrix}, ~~
   \mathbf{C}_e=
   \begin{bmatrix}
   \mathbf{0}&\mathbf{C}_2
   \end{bmatrix},
\end{align}
where $\mathbf{A}_{12}\in \mathbb{R}^{1\times(2n-1)}$ and $\mathbf{A}_2\in \mathbb{R}^{(2n-1)\times (2n-1)}$, $\mathbf{B}_2\in\mathbb{R}^{(2n-1)\times 2n}$ and $\mathbf{C}_2$ is the matrix obtained by removing the 
first column of the  matrix $\mathbf{C}_e$ so that
\begin{align}\label{C2-delta-omega}
\mathbf{C}_2=
\begin{bmatrix}
\widetilde{\mathbf{C}}^{\bm\top}\mathbf{M}^{-1/2}\widehat{\mathbf{U}}&\bm 0\\
\mathbf{0} &\mathbf{M}^{-1/2}\mathbf{U}
\end{bmatrix}
\in\mathbb{R}^{(m+n)\times (2n-1)},
\end{align}
with $\widehat{\mathbf{U}}=
\begin{bmatrix}
\mathbf{u}_2&\mathbf{u}_3&\cdots&\mathbf{u}_n
\end{bmatrix}
\in\mathbb{R}^{n\times (n-1)}$. 
The variance matrix $\mathbf{Q}_y$ of the output $\mathbf{y}$ of the system (\ref{stochasticsystem}) in the invariant probability distribution  satisfies
\begin{align}\label{Qy}
\mathbf{Q}_y=\mathbf{C}_2\mathbf{Q}_x\mathbf{C}_2^{\bm\top}
\end{align}
where
$\mathbf{Q}_x\in\mathbb{R}^{(2n-1)\times (2n-1)}$ is the unique solution of the following Lyapunov equation
\begin{align}\label{barQ}
\mathbf{A}_2\mathbf{Q}_x+\mathbf{Q}_x\mathbf{A}_2^{\bm\top}+\mathbf{B}_2\mathbf{B}_2^{\top}=\mathbf{0}
\end{align}
\end{theorem}
\par 
With the assumption of the uniform disturbance-damping ratio $b_i^2/d_i$ at all the nodes, i.e., $b_i^2/d_i=b_j^2/d_j$
for $i,j\in\mathcal{V}$, the explicit formula the $\mathbf{Q}$ have been deduced
in \cite{ZhenWangAutom}, from which the role of the network topology is revealed. 
However, the propagation of the fluctuations cannot be fully illustrated with this assumption. 
\par
To emphasize the effect of the inertia in the system (\ref{stochasticsystem}), we also study the fluctuations in the stochastic process
\begin{subequations}\label{kuramoto_linear}
\begin{align}
        \text{d}\overline{\bm\delta}(t)&=-\mathbf{D}^{-1}\mathbf{L}\overline{\bm\delta}(t)\text{d}t+\mathbf{D}^{-1}\widetilde{\mathbf{B}}\text{d}\bm v(t),\\
        \overline{\bm y}(t)&=\widetilde{\mathbf{C}}^{\top}\overline{\bm\delta}(t),
\end{align}
\end{subequations}
which is the linearization of the non-uniform Kuramoto model \cite{Dorfler2012,Kaihua_SIAM}.  
This system can also be obtained by setting $m_i=0$ in the system (\ref{stochasticsystem}) at all the nodes. 
Denote the matrix $\overline{\mathbf{U}}\in\mathbb{R}^{n\times n}$ such that 
\begin{eqnarray}
\overline{\mathbf{U}}^{\top}\mathbf{D}^{-1/2}
            \mathbf{L}
          \mathbf{D}^{-1/2} \overline{\mathbf{U}}&=&\overline{ \mathbf{\Lambda}}_n \label{diagonal} 
\end{eqnarray}
where $\overline{ \mathbf{\Lambda}}_n=(\overline{\lambda}_i)\in\mathbb{R}^{n\times n}$ with 
$\overline{\lambda}_i$ being the eigenvalue of the matrix $\mathbf{D}^{-1/2}
            \mathbf{L}\mathbf{D}^{-1/2}$.  The matrix $\overline{\mathbf{U}}$ is further decomposed
 into the form $
 \overline{\mathbf{U}}=\begin{bmatrix}
 \overline{\mathbf{u}}_1&\overline{\mathbf{U}}_2
 \end{bmatrix}$

 For the model (\ref{kuramoto_linear}),  the variance matrix of the phase 
difference is presented in the following theorem \cite{Kaihua_SIAM}.
\begin{theorem}\label{theoremmain_kuramoto}
Consider the stochastic system (\ref{kuramoto_linear}) with a connected graph $\mathcal{G}$. 
The  asymptotic variance of the output process $\overline{\mathbf{y}}$ can be 
computed by 
\begin{eqnarray}
  \overline{\mathbf{Q}}_\delta
    & = &\widetilde{\mathbf{C}}^{\top}\mathbf{D}^{-1/2}
              \overline{\mathbf{U}}_2
                \overline{\mathbf{Q}}_{x}
             \overline{\mathbf{U}}_2^{\top}\mathbf{D}^{-1/2}
          \widetilde{\mathbf{C}}.  \label{Qdelta} 
\end{eqnarray}
where $\overline{\mathbf{U}}_2 
    = \begin{bmatrix}
     \overline{\mathbf{u}}_2 & \overline{\mathbf{u}}_3 & \ldots & \overline{\mathbf{u}}_{n}
    \end{bmatrix}\in \mathbb{R}^{n \times (n-1)}$ and $\overline{\mathbf{Q}}_{x} = (\overline{q}_{x_{i,j}}) \in \mathbb{R}_{spd}^{(n-1)\times (n-1)}$ is the unique solution of the Lyapunov equation,
\begin{eqnarray}\label{LyapunovQnplus1}
\mathbf{0}
    & = & -\overline{ \mathbf{\Lambda}}_{n-1} \overline{\mathbf{Q}}_{x} 
          - \overline{\mathbf{Q}}_{x} \overline{\mathbf{\Lambda}}_{n-1} 
          + \overline{\mathbf{U}}_2^\top\mathbf{D}^{-1/2}\widetilde{\mathbf{B}}\widetilde{\mathbf{B}}^\top\mathbf{D}^{-1/2}\overline{\mathbf{U}}_2, ~
\end{eqnarray}
with $\overline{\mathbf{\Lambda}}_{n-1}
          = \text{diag}( \overline{\lambda}_2, ~ \overline{\lambda}_3, ~ \ldots, ~ \overline{\lambda}_n )\in\mathbb{R}_{diag}^{(n-1) \times (n-1)}$.
In addition, the matrix $\overline{\mathbf{Q}}_{x}$ is solved from the Lyapunov equation as
\begin{align}
  \!\overline{q}_{x_{i,j}}
    & = & (\overline{\lambda}_{i+1}+\overline{\lambda}_{j+1})^{-1}
         \overline{\mathbf{u}}_{i+1}^{\top}\mathbf{D}^{-1/2}
          \widetilde{\mathbf{B}}\widetilde{\mathbf{B}}^{\top}\mathbf{D}^{-1/2}
          \overline{\mathbf{u}}_{j+1},  \forall ~ i, ~ j = 1, ~ \cdots, ~ n-1, \!\label{Q1}
\end{align}
and in particular, 
\begin{eqnarray}
        \overline{q}_{x_{i,i}}
    & = & \frac{1}{2} \overline{\lambda}_{i+1}^{-1}
          \overline{\mathbf{u}}_{i+1}^{\top}
            \mathbf{D}^{-1/2}
              \widetilde{\mathbf{B}}\widetilde{\mathbf{B}}^{\top}
            \mathbf{D}^{-1/2}
          \overline{\mathbf{u}}_{i+1},~
          \forall ~ i = 1, ~ \cdots, ~ n-1. \label{Q2}
\end{eqnarray}

\end{theorem}

\par 
\section{Main results}\label{main:results}
In this section, we present the main results of this paper. The reader may find the proofs of the results
in Section~\ref{Section:proofs}. 
We focus on multi-machine systems (\ref{stochasticsystem}).  Based on the following assumption, we derive the explicit formula
of the solution $\mathbf{Q}_y$. 
\begin{assumption}\label{assumption}
Consider the stochastic system (\ref{stochasticsystem}), assume the damping-inertia ratios $d_i/m_i $ are uniform at
all the nodes, i.e., for all $i\in\mathcal{V}$, $d_i/m_i=\alpha\in\mathbb{R}_+$. 
\end{assumption}
However, in practice the differences of the ratios$d_i/m_i$ are relatively small because the inertia and the damping 
are usually proportional to the rating of the power generators. Assumption \ref{assumption} allows us to derive explicit formulas to reveal the propagation of the fluctuations in the networks. 
Following Theorem \ref{theoremmain0}, we obtain the following theorem.
\par 
\begin{theorem}\label{maintheorem_kuramot}
Consider the invariant probability distribution of the system (\ref{stochasticsystem}). 
Decompose the matrix $\mathbf{Q}_x$ defined in Theorem \ref{theoremmain0} into matrices, 
\begin{align}\label{equationQx}
\mathbf{Q}_x=
\begin{bmatrix}
\mathbf{G}&\mathbf{S}\\
\mathbf{S}^{\bm\top}&\mathbf{R}
\end{bmatrix}
\end{align}
where $\mathbf{G}=(g_{i,j})\in\mathbb{R}^{(n-1)\times(n-1)}$ which satisfies $\mathbf{G}=\mathbf{G}^\top$
, $\mathbf{S}=(s_{i,j})\in\mathbb{R}^{(n-1)\times n}$ and $\mathbf{R}=(r_{i,j})\in\mathbb{R}^{n\times n}$ which satisfies $\mathbf{R}=\mathbf{R}^\top$. 
The variance matrix $\mathbf{Q}_y$ with the form of block matrix in (\ref{Q_y_form}) satisfies 
\begin{subequations}\label{eq:variance0}
    \begin{align}
        \mathbf{Q}_\delta&=\widetilde{\mathbf{C}}^\top\mathbf{M}^{-1/2}\widehat{\mathbf{U}}\mathbf{G}\widehat{\mathbf{U}}^{ \top}\mathbf{M}^{-1/2}\widetilde{\mathbf{C}},\\
        \mathbf{Q}_\omega&=\mathbf{M}^{-1/2}\mathbf{U}\mathbf{R}\mathbf{U}^{\bm\top}\mathbf{M}^{-1/2},\\
        \mathbf{Q}_{\delta\omega}&=\mathbf{M}^{-1/2}\mathbf{U}\mathbf{S}^{\top}\widehat{\mathbf{U}}^{\bm \top}\mathbf{M}^{-1/2}\widetilde{\mathbf{C}}.
    \end{align}
\end{subequations}
Define 
\begin{align*}
\rho_i=2\alpha^2+\lambda_{i},~~\chi_{i,j}=(\lambda_{i}-\lambda_{j})^2+2\alpha^2(\lambda_{j}+\lambda_{i}).
\end{align*}
If Assumption \ref{assumption} holds, then $\mathbf{Q}_y$ can be solved from (\ref{eq:variance0})
with explicit formula of $\mathbf{Q}_x$ solved from the Lyapunov equation (\ref{barQ}), where 
$\mathbf{S}$ satisfies
for $i=1,2,\cdots,n-1$, 
 \begin{align}\label{eq:result1}
  s_{i,1}=&\rho_{i+1}^{-1}\mathbf{u}_{i+1}^{\top} \mathbf{M}^{-1/2}\widetilde{\mathbf{B}}^2\mathbf{M}^{-1/2}\mathbf{u}_{1},
\end{align}
for $i,j=2,3,\cdots,n$, 
\begin{align}\label{eq:result2}
s_{i-1,j}=&\frac{\lambda_{i}-\lambda_{j}}{\chi_{i,j}}
                {\mathbf{u}}_{i}^{\top}{\mathbf{M}}^{-1/2}\widetilde{\mathbf{B}}^2\mathbf{M}^{-1/2}\mathbf{u}_{j};
\end{align}
$\mathbf{G}$ satisfies for $i,j=2,3,\cdots,n$,
\begin{align}\label{eq:result3}
 g_{i-1,j-1}&=\frac{2\alpha}{\chi_{i,j}}\mathbf{u}_{i}^{\top}
 \mathbf{M}^{-1/2}\widetilde{\mathbf{B}}^2\mathbf{M}^{-1/2}\mathbf{u}_{j};      
\end{align}
$\mathbf{R}$ satisfies 
\begin{align}\label{eq:result4}
    r_{1,1}=\frac{1}{2\alpha}
    \mathbf{u}_{1}^{\top}
     \mathbf{M}^{-1/2}\widetilde{\mathbf{B}}^2\mathbf{M}^{-1/2}\mathbf{u}_{1}.
\end{align}
for $i,j=1,2,\cdots,n,$ with $(i,j)\neq (1,1)$,
\begin{align}\label{eq:result5}
r_{i,j}=\frac{\alpha(\lambda_{i}+\lambda_{j})}{\chi_{i,j}}
\mathbf{u}_{i}^{\top}
 \mathbf{M}^{-1/2}\widetilde{\mathbf{B}}^2\mathbf{M}^{-1/2}\mathbf{u}_{j}.
\end{align}
Here $\widetilde{\mathbf{B}}^2=\widetilde{\mathbf{B}}\widetilde{\mathbf{B}}^\top$ 
because $\widetilde{\mathbf{B}}$ is a diagonal matrix. 
\end{theorem}
\par 
See Section \ref{Section:proofs} for the proof of this theorem.  Following this theorem, 
it is found that the impact of the disturbances can be 
described by the \emph{Superposition Principle}.  This property demonstrates that 
the fluctuations in the system caused by the disturbance at a node can never 
be balanced by the disturbances at the other nodes. 
\par 
To reveal the influences of the system parameters on the 
fluctuations more explicitly,  we further make an assumption as follows.
\begin{assumption}\label{assumption2}
Assume that the inertia and the damping of the synchronous machines are all identical in the system, i.e., $\mathbf{M}=\eta\mathbf{I}_n$ and $\mathbf{D}=d\mathbf{I}_n$, which leads to $\alpha=d/\eta$. 
\end{assumption}
\par 
Clearly, this assumption is more restrictive than Assumption \ref{assumption}, with which
we obtain the following corollary for the trace of the variance matrix of the frequency (\ref{eq:result4}). 
\begin{corollary}\label{corollary-trace}
Consider the system (\ref{stochasticsystem}). If Assumption \ref{assumption2} holds, 
then the variance matrix of the frequency satisfies, 
\begin{align*}
\emph{tr}(\mathbf{Q}_\omega)=\frac{1}{2d\eta}\emph{tr}(\widetilde{\mathbf{B}}^2). 
\end{align*}
\end{corollary}
\par 
The proof follows immediately from $\text{tr}(\mathbf{R} )=\frac{1}{2\alpha\eta}\text{tr}(\widetilde{\mathbf{B}}^2)$ 
with the fact that pre- and post-multiplication of the matrix
$\widetilde{\mathbf{B}}^2$ by the orthogonal matrix $\mathbf{U}$
according to $\mathbf{U} \widetilde{\mathbf{B}}^2 \mathbf{U}^\top$ will not change the trace of this matrix.
Following from this corollary, it is found that adding new nodes without any disturbances will not change
the total amount of fluctuations in the network if Assumption \ref{assumption2} is satisfied. 
It is shown that the trace of the variance
matrix of the frequency is independent on the network
topology. However, it will be shown in the next section
that the variance of the frequency at each node depend
on the network topology. 
\par 
Based on Assumption \ref{assumption2} and Theorem \ref{maintheorem_kuramot}, we investigate the propagation 
of the disturbance in two types of special graphs, i.e., complete graphs and star graphs.
For simplicity, we further make an assumption on the 
weight of the lines as below. 
\begin{assumption}\label{assumption-uniformlines}
Assume the weights of the lines in the graph are all identical, i.e., $K_{i,j}\cos\delta_{ij}^*=\gamma$ for $(i,j)\in\mathcal{E}$.
\end{assumption}
This assumption is practical for the power networks with identical line capacities and small phase
 differences at the synchronous state., i.e., $\delta_i^*\approx\delta_j$ for all $(i,j)\in\mathcal{E}$. 
It allows us to deduce the explicit formulas of 
the variance matrix of the frequency and the phase differences in the power systems
with complete graphs and star graphs. 

\subsection{Complete graphs}
For a power systems with a complete graph,  it yields the following proposition from Theorem \ref{maintheorem_kuramot} and Lemma \ref{lemma_eigenvalues}. 
\begin{proposition}\label{Corollary-complete}
Consider the system (\ref{stochasticsystem}) with a complete graph. If
Assumption \ref{assumption2} and \ref{assumption-uniformlines} holds, then the variance of the frequency at node $i$ for $i=1,2,\cdots,n$ satisfies
\begin{align}\label{completenetork1}
       q_{\omega_{i,i}}
        =\big[\frac{1}{2d\eta}-\frac{\gamma(n-1)}{dn(2d^2+\gamma\eta n)}\big]b_i^2+\frac{\gamma}{d n\left(2d^2+\gamma \eta n\right)}(\emph{tr}(\widetilde{\mathbf{B}}^2)-b_{i}^2)
    \end{align}
and the variance matrix  $\bm Q_\delta$ of the phase angle difference satisfies
        \begin{equation}\label{completenetowrk2}
            \mathbf{Q}_\delta=\frac{1}{2d\gamma  n}{\widetilde{\mathbf{C}}}^{\top}\widetilde{\mathbf{B}}^2\widetilde{\mathbf{C}}.
        \end{equation}
      In particular,  for the line $e_k$ connecting node $i$ and $j$, the variance of the phase angle difference in this line is 
      \begin{align}
      q_{\delta_{k,k}}=\frac{1}{2d\gamma n}(b_i^2+b_j^2), \label{eq:Q_delta}
      \end{align}
     and the trace of $\mathbf{Q}_\delta$ satisfies
\begin{equation}\label{eq:star3}
    \emph{tr}(\mathbf{Q}_\delta)=\frac{n-1}{2d\gamma n}\emph{tr}(\widetilde{\mathbf{B}}^2).
\end{equation}
\end{proposition}
\par 
The next corollary of Proposition \ref{Corollary-complete} explains the finding on the propagation of the fluctuations from a node to the others in details.
\begin{corollary}\label{corollay-complete_graph}
Consider the system (\ref{stochasticsystem}) with a complete graph. If
Assumption \ref{assumption2} and \ref{assumption-uniformlines} holds, and $b_i\neq 0$ and $b_j=0$ for all $j$ with $j\neq i$, then
\begin{align}
q_{\omega_{i,i}}&= \frac{b_i^2}{2d\eta}-\frac{(n-1)\gamma b_i^2}{dn(2d^2+\gamma\eta n)},\label{Q_omega_i}\\
q_{\omega_{j,j}}&= \frac{\gamma b_{i}^2}{d n\left(2d^2+\gamma \eta n\right)},~\text{for}~ j\neq i,\label{Q_omega_j}
\end{align}
and the variances of the phase angle differences satisfy
\begin{eqnarray}
    q_{\delta_{k,k}}
    & = & \left\{
          \begin{array}{ll}
           \frac{b_i^2}{2d\gamma n} & \mbox{if line $e_k$ is connected to node $i$} \\
            0,               & \mbox{else.}
          \end{array}
          \right.
\end{eqnarray}
\end{corollary}
\par 
For comparison, the asymptotic matrix of the phase differences in the model (\ref{kuramoto_linear}) 
is presented in the following proposition with proof in Section \ref{Section:proofs}. 
\begin{proposition}\label{proposition47}
 Consider the system (\ref{kuramoto_linear}) with a complete graph. Assume $\mathbf{D}=d\mathbf{I}$ and 
Assumption \ref{assumption-uniformlines} holds, then the variances of the phase angle differences satisfy
        \begin{equation}\label{complete_kuramoto2}
           \overline{\mathbf{Q}}_\delta=\frac{1}{2d\gamma  n}{\widetilde{\mathbf{C}}}^{\bm\top}\widetilde{\mathbf{B}}^2\widetilde{\mathbf{C}},
                   \end{equation}
       with 
      \begin{align}
      \overline{q}_{\delta_{k,k}}=\frac{1}{2d\gamma n}(b_i^2+b_j^2),~\text{for}~k=1,\cdots,m.\label{eq:bar_Q_delta}
      \end{align}
\end{proposition}
\par
\begin{figure}\label{Fig_complete}
    \centering
    \includegraphics[scale=1.6]{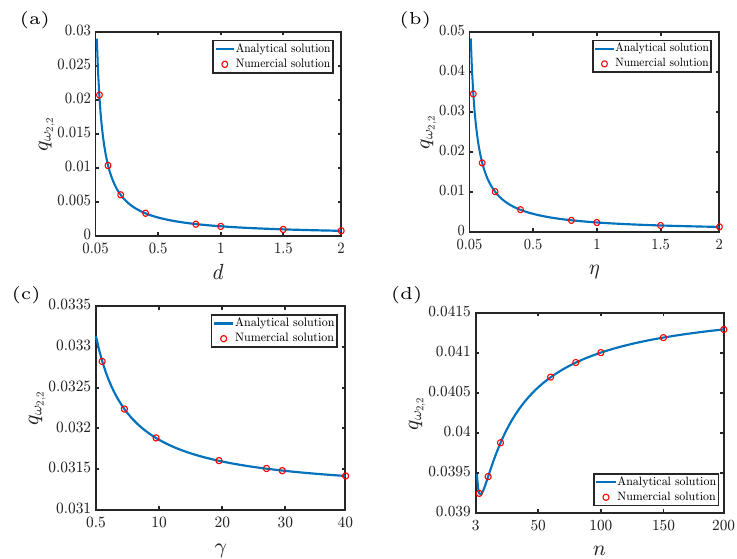}
    \caption{The dependence of the variance $q_{\omega_{2,2}}$ on the system parameters in the complete graph with $b_2\neq 0$ and $b_i=0$ for $i\neq 2$ and indices of the nodes and lines defined in Definition \ref{definition_index}(a).}
    \end{figure}

\begin{figure}\label{Fig_complete_phase}
    \centering
    \includegraphics[scale=1.6]{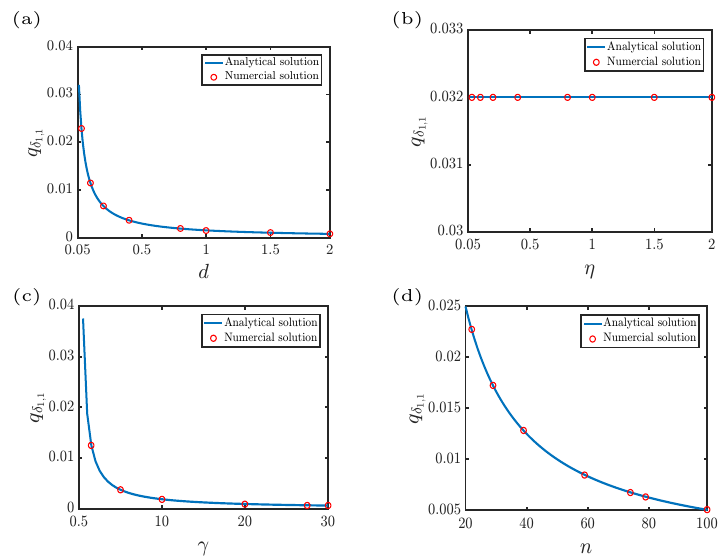}
    \caption{The relationship for the variance $q_{\delta_{1,1}}$ for line $(1,2)$ in the complete graph with $b_2\neq 0$ and $b_i=0$ for $i\neq 2$ and indices of the nodes and lines defined in Definition \ref{definition_index}(a).}
    \end{figure}

 
To verify the correctness of these analytical formulas in Corollary \ref{proposition47}, we use Matlab to compute the variances 
numerically from (\ref{Qy}) and (\ref{barQ}) in the complete graph with $b_2\neq 0$ and $b_i=0$ for $i\neq 0$ and 
indices of the nodes and lines defined in Definition \ref{definition_index}(a). In order to satisfy
Assumption \ref{assumption-uniformlines}, we set $P_i=0$ for all the nodes and $K_{ij}=K$ for all 
the lines, which leads to $\delta_i^*-\delta_j^*=0$ at the 
synchronous state and the line weight $\gamma=K\cos{\delta_{ij}^*}=K$. 
The setting of the parameters for plotting these figures are shown in Table \ref{table_complete}. 
It is shown in Fig. \ref{Fig_complete} and Fig. \ref{Fig_complete_phase} that the analytical solution and the numerical solution of the variances are all identical.

\begin{table*}[htbp]
  \begin{center}
      \caption{The setting of the parameters for plotting Fig.\ref{Fig_complete} and \ref{Fig_complete_phase}.}
      \label{table_complete}
      \scalebox{0.9}{
      \begin{tabular}{|c|c|c|c|c||c|c|c|c|}
      \hline
      \multirow{2}*{parameters}&\multicolumn{4}{c||}{Fig.\ref{Fig_complete}}&\multicolumn{4}{c|}{Fig.\ref{Fig_complete_phase}}\\
          \cline{2-9}
        &(a)&(b)&(c)&(d)&(a)&(b)&(c)&(d)\\
          \hline
          $\gamma$ &10&10&$-$&10&10&10&$-$&10\\
          \hline
          $\eta$ &0.5&$-$&0.02&0.02&0.5&$-$&0.01&0.1\\
          \hline
          $d$ &$-$&0.3&1.2&1.5&$-$&0.1&1.2&0.1\\
           \hline
          $b_2$ &0.04&0.04&0.04&0.05&0.8&0.8&1.5&1\\ 
                \hline
          $n$ &20&20&30&$-$&20&10&50&$-$\\   
          \hline           
      \end{tabular}
      }				
  \end{center}
\end{table*}

\par 
Based on Corollary \ref{corollay-complete_graph} and Proposition \ref{proposition47}, we get the following findings on the variance of the frequency 
and the phase differences in the stochastic system (\ref{kuramoto_linear}) with the complete graph. 
\par 
{\bf (a) On the variance of the frequency in the complete graph.}
As either the inertia $\eta$ or the damping $d$ of the synchronous machines increases, the variance of 
the frequencies at all the nodes decrease. This statement is well known to experts in the field
and will not be discussed further. 
There are two terms in the right hand side of (\ref{Q_omega_i}), in which the first term is the variance of the fluctuations introduced by introduced by the disturbance at node $i$ and the second term measures the fluctuations 
propagated from node $i$ to all the other nodes. Thus, we only need to analyze the dependence of the variance of the 
frequency at node $i$
on the weight of the lines and the network size. 
\par 
First, we introduce the impact of the line weights. On contrary to the case of SMIB model, the weights of the lines play roles on the variance 
of the frequency. The derivative of the variance with respect 
to $\gamma$ satisfy
\begin{align*}
\frac{\partial q_{\omega_{i,i}}}{\partial \gamma}=\frac{2d(1-n)}{ n(2d^2+\gamma \eta n)^2}b_i^2<0,
~~\text{and} ~~\frac{\partial q_{\omega_{j,j}}}{\partial \gamma}=\frac{2d}{ n(2d^2+\gamma \eta n)^2}b_i^2>0,
\end{align*} 
This indicates that, as the line weights increase, the frequency variance at the source node
of the disturbance decreases while those at the other nodes increase. Thus, increasing the line capacities, which increases the line weights, will increase the propagate of the fluctuation from the source node to the other nodes. 
However, there exists a lower bound for the variance of the frequency at nodes and an upper bound 
for the variance of the frequency at the other nodes, which are the limits of the variance as $\gamma$ goes to infinity
respectively,
\begin{align*}
\lim_{\gamma\rightarrow \infty}q_{\omega_{i,i}}=\frac{1}{2d\eta}b_i^2-\frac{n-1}{d\eta n^2}b_i^2,
~~\text{and}~~
\lim_{\gamma\rightarrow \infty}q_{\omega_{j,j}}=\frac{1}{d\eta n^2}b_i^2. 
\end{align*}
\par 
Second, we focus on the impact of the network size. 
From (\ref{Q_omega_i}), it yields
\begin{align*}
\lim_{n\rightarrow \infty}q_{\omega_{i,i}}=\frac{b_i^2}{2d\eta}.
\end{align*}
Clearly, this limit equals to the value of the frequency variance presented in (\ref{SMIBQ}) for the SMIB model. 
This indicates that the network becomes an infinite bus for node $i$ when the size is sufficiently large. 
If the size of the network is large, then it holds
\begin{align*}
\frac{1}{2d\eta}b_i^2\gg\frac{\gamma(n-1)}{dn(2d^2+\gamma\eta n)} b_i^2
\end{align*}
which demonstrates that the disturbance impacts the local node most. 
In addition, the derivative of the variances with respect to $n$ satisfies
\begin{align*}
\frac{\partial q_{\omega_{i,i}}}{\partial n}&=\frac{\gamma(\gamma\eta n^2-2\gamma\eta n-2d^2)}{d(2d^2n+\gamma\eta n^2)^2}b_i^2,~~\text{and}~~ \frac{\partial q_{\omega_{j,j}}}{\partial n}=\frac{-\gamma(2d^2+2\gamma\eta n)}{d(2d^2 n+\gamma\eta n^2)^2}b_i^2<0. 
\end{align*}
It is found that if $n>n_c$ with $n_c=\lfloor1+\sqrt{1+\frac{2d^2}{\gamma\eta}}\rfloor$ defined 
as a critical value of the network size, then
$$\frac{\partial q_{\omega_{i,i}}}{\partial n}>0.$$
This indicates that the variance of the frequency at node $i$ 
increases as the size of the network increases.  This trend 
is shown in Fig. \ref{Fig_complete}(d) for the graph with $b_2\neq 0$ and $b_j=0$ for $j\neq 2$. 

\emph{It is found that when $n>n_c$, increasing 
the size of the network have a negative impact on suppressing the frequency variance at node $i$.  In other words,
adding new nodes to the network prevents the propagation of the fluctuations from node $i$ to the other nodes.} 
In addition, for any $n\geq 2$, it holds
\begin{align*}
q_{\omega_{i,i}}\geq \big(\frac{1}{2d\eta}-\frac{\gamma}{d(\sqrt{\gamma\eta}+\sqrt{\gamma\eta+2d^2})^2}\big)b_i^2. 
\end{align*}
which shows the lower bound of the variance of the frequency at node $i$. 
\par 
{\bf (b) On the variance of the phase difference in a complete graph.}
The roles of the damping coefficient $d$, the line weight $\gamma$, the graph size $n$ 
can be clearly seen from the formula (\ref{eq:Q_delta}). 
Because the inertia $\eta$ is absent in this formula, the variance is independent on the inertia of 
the node. Due to this independence, the variance matrix of the phase difference
in the system (\ref{stochasticsystem}) and the system (\ref{kuramoto_linear}) are equal, i.e., $\overline{\mathbf{Q}}_\delta=\mathbf{Q}_\delta$, which 
is verified by the formula (\ref{completenetowrk2}) and (\ref{complete_kuramoto2}). It is surprisingly found that the variance only depends on the disturbance
from the node $i$ and $j$ while it is independent on the disturbances from all the other nodes. 
In addition, as the size of the network increases, the variances of the 
phase angle differences in the lines connecting node $i$ decreases. This is because 
as the size of the complete graph increases, the lines connecting node $i$ also increases, which 
share the fluctuation from node $i$.  
\par

\subsection{Star graphs}
In this subsection, we study the variance matrices in the systems with star graphs.  Based 
on Theorem \ref{maintheorem_kuramot} and Lemma \ref{lemma_eigenvalues}, we obtain 
the following result. 
\begin{proposition}\label{Corollary-star}
Consider the system (\ref{stochasticsystem}) with a star graph where 
the indices of the nodes and lines are defined as in Definition \ref{definition_index}(ii).    
If
the Assumption \ref{assumption2} and \ref{assumption-uniformlines} both holds, then the variance 
$\mathbf{Q}_\omega$ of the frequency satisfies
    \begin{align}\label{eq:star1}
 q_{\omega_{1,1}}=\big[\frac{1}{2d\eta}-\frac{\gamma(n-1)}{dn(2d^2+\gamma\eta n)}\big]b_{1}^2+\frac{\gamma}{d n\left(2d^2+\gamma \eta n\right)}(\emph{tr}(\widetilde{\mathbf{B}}^2)-b_{1}^2)
        \end{align}
    and for $i=2,3,\cdots,n$,
    \begin{equation*}
        \begin{aligned}
        q_{\omega_{i,i}}
        =&
       \frac{\gamma b_{1}^2}{d n\left(2d^2+\gamma \eta n\right)}+\frac{b_i^2}{2d\eta}-\frac{\gamma b_i^2}{dn(2d^2+\gamma\eta n)}-\frac{\gamma(n-2)b_i^2}{dn(2d^2(n+1)+\gamma\eta(n-1)^2)}\\
       &-\frac{\gamma^2\eta(n-2)b_i^2}{dn(2d^2+\gamma\eta)(2d^2+\gamma\eta n)}+\frac{\gamma (\emph{tr}(\widetilde{\mathbf{B}}^2)\!-\!b_i^2\!-\!b_1^2)}{dn(2d^2(1\!+\!n)\!+\!\gamma\eta (n-1)^2)}\\
        &+\frac{\gamma^2\eta}{dn(2d^2\!+\!\gamma\eta)(2d^2\!+\!\gamma\eta n)}(\emph{tr}(\widetilde{\mathbf{B}}^2)\!-\!b_i^2\!-\!b_1^2)
        \end{aligned}
    \end{equation*}
and the variance matrix $\mathbf{Q}_\delta$ of the phase angle differences satisfies
for $k\neq q$,
        \begin{equation*}
                \begin{aligned}
                    q_{\delta_{k,q}}
                    =&\frac{2d^2(n+1)+\gamma\eta(n-1)^2}{2d\gamma n(2d^2(1+n)+\gamma\eta(n-1)^2)} b_{1}^2+ \frac{-2d^2(n-1)+\gamma\eta(2n-n^2+1)}{2d\gamma n(2d^2(1+n)+\gamma\eta(n-1)^2)} b_{k+1}^2\\
                    &+
                    \frac{-2d^2(n-1)+\gamma\eta(2n-n^2+1)}{2d\gamma n(2d^2(1+n)+\gamma\eta(n-1)^2)} b_{q+1}^2\\
                    &+\frac{(2d^2+\gamma\eta(n+1))\left(\emph{tr}(\widetilde{\mathbf{B}}^2)-b_{k+1}^2-b_{q+1}^2-b_1^2\right)}{2d\gamma n(2d^2(1+n)+\gamma\eta(n-1)^2)} 
             \end{aligned}
        \end{equation*}
        and for $k=1,\cdots,m$, 
        \begin{equation}\label{eq:star2}
            \begin{aligned}
              q_{\delta_{k,k}}
               =&\frac{1}{2d\gamma n} b_{1}^2+\big(\frac{n-1}{2d\gamma n}-
               \frac{(n-2)(2d^2+\gamma\eta(n+1))}{2d\gamma n(2d^2(1+n)+\gamma\eta(n-1)^2)}\big) b_{k+1}^2\\\
               &+\frac{(2d^2+\gamma\eta(n+1)) \left(\emph{tr}(\widetilde{\mathbf{B}}^2)-b_{k+1}^2-b_1^2\right)}{2d\gamma n(2d^2(1+n)+\gamma\eta(n-1)^2)}
         \end{aligned}
    \end{equation}
    and the trace of $\mathbf{Q}_\delta$ satisfies
\begin{equation}\label{eq:star4}
    \emph{tr}(\mathbf{Q}_\delta)=\frac{n-1}{2d\gamma n}\emph{tr}(\widetilde{\mathbf{B}}^2).
\end{equation}
\end{proposition}
\par 
See the proof of this proposition in Section \ref{Section:proofs}.
With these explicit formulas, we investigate the propagation 
of the disturbances in the star graphs. We first focus on 
the graphs with a disturbance at the root node and then 
on the networks with a disturbance at a non-root node.
\par 
\begin{corollary}\label{corollary-stargraph-1}
Consider the system (\ref{stochasticsystem}) with a star graph where 
the indices of the nodes and lines are defined as in Definition \ref{definition_index}(ii).    
If
Assumption \ref{assumption2} and \ref{assumption-uniformlines} holds and there are disturbances at the 
root node $i=1$
and no disturbances at all the other nodes, i.e., $b_1\neq 0$ and $b_i=0$ for $i=2,\cdots,n$, then
the variances matrix $\mathbf{Q}_\omega$ of the frequencies satisfies
\begin{align*}
 q_{\omega_{1,1}}=\big[\frac{1}{2d\eta}-\frac{\gamma(n-1)}{dn(2d^2+\gamma\eta n)}\big]b_{1}^2,
\end{align*}
and for the other nodes, 
\begin{align*}
q_{\omega_{i,i}}= \frac{\gamma}{d n\left(2d^2+\gamma \eta n\right)}b_{1}^2,~i=2,\cdots,n,
\end{align*}
and the variances  $\mathbf{Q}_\delta$ of the phase angle differences satisfy
\begin{align*}
q_{\delta_{k,k}}=\frac{1}{2d\gamma n} b_{1}^2,~k=1,\cdots,n-1.
\end{align*}
\end{corollary}
\par 
It is clearly seen in this corollary that the formulas are all the same to the ones in Corollary 
\ref{corollay-complete_graph} when $i=1$. This demonstrates that when there are disturbance at the 
root node $i=1$ only in the star graph, the dependence of the variances of the frequency and 
the phase difference on the system parameters, i.e., the synchronous machines' inertia and 
damping, the size of the network and the weights of the lines, are total the same as in 
the complete graph, which will not be explained again. 
\par 
If the disturbances occurs at the non-root node, we obtain the following corollary. 
\par 
\begin{corollary}\label{corollary-stargraph_2}
Consider the system (\ref{stochasticsystem}) with a star graph where 
the indices of the nodes and lines are defined as in Definition \ref{definition_index}(ii).   
If
Assumption \ref{assumption2} and \ref{assumption-uniformlines} holds and there are disturbances at node $i=2$ and no disturbances at all the other nodes,
i.e., $b_2\neq 0$ and $b_1=0$ and $b_i=0$ for $i=3,\cdots,n$, then the variances matrix $\mathbf{Q}_\omega$ 
of the frequencies satisfies, 
\begin{align}
q_{\omega_{1,1}}&=\frac{\gamma}{dn(2d^2+\gamma \eta n)}b_2^2,\\
q_{\omega_{2,2}}&=\frac{b_2^2}{2d\eta}-\frac{\gamma b_2^2}{dn(2d^2+\gamma\eta n)}-\frac{\gamma(n-2)b_2^2}{dn(2d^2(n+1)+\gamma\eta(n-1)^2)} \label{StarQ-omega}\\
\nonumber
&-\frac{\gamma^2\eta(n-2)b_2^2}{dn(2d^2+\gamma\eta)(2d^2+\gamma\eta n)},\\
\nonumber
\text{and for}&~~i=3,\cdots,n,\\
q_{\omega_{i,i}}&=\frac{\gamma}{dn(2d^2(1\!+\!n)\!+\!\gamma\eta (n-1)^2)}b_2^2+\frac{\gamma^2\eta}{dn(2d^2\!+\!\gamma\eta)(2d^2\!+\!\gamma\eta n)}b_2^2,
\end{align}
the variances matrix $\mathbf{Q}_\delta$ of the phase differences satisfies, 
\begin{align}
q_{\delta_{1,1}}&=\big(\frac{n-1}{2d\gamma n}-
               \frac{(n-2)(2d^2+\gamma\eta(n+1))}{2d\gamma n(2d^2(1+n)+\gamma\eta(n-1)^2)}\big) b_{2}^2,\label{starQ_phase1}\\
q_{\delta_{k,k}}&=\frac{2d^2+\gamma\eta(n+1)}{2d\gamma n(2d^2(1+n)+\gamma\eta(n-1)^2)} b_{2}^2.\label{starQ_phase2}
\end{align}
\end{corollary}
To emphasize the impact of the inertia, we deduce the variance matrix of the system (\ref{kuramoto_linear}) with a star graph.
\begin{proposition}\label{proposition_kuramoto}
Consider the system (\ref{kuramoto_linear}) with a star graph where 
the indices of the nodes and lines are defined as in Definition \ref{definition_index}(ii).      
Assume $\mathbf{D}=d\mathbf{I}_n$ and Assumption \ref{assumption-uniformlines} holds, then
the matrix $\overline{\mathbf{Q}}_\delta$ satisfies,
    \begin{align}
       \!\overline{q}_{\delta_{k,q}}&=\frac{b_1^2}{2d\gamma n}+\frac{(1-n)(b_{k+1}^2+b_{q+1}^2)}{2d\gamma n(1+n)}+\frac{1}{2d\gamma n(1+n)}\left(\text{tr}\left(\widetilde{\mathbf{B}}^2\right)-b_{k+1}^2-b_{q+1}^2-b_1^2\right)\!,\label{overlineQ0}
    \end{align} 
    and 
    \begin{align}
        \overline{q}_{\delta_{k,k}}=\frac{1}{2d\gamma n}b_1^2+\frac{n^2-n+1}{2d\gamma n(1+n)}b_{k+1}^2+\frac{1}{2d\gamma n(1+n)}\left(\text{tr}\left(\widetilde{\mathbf{B}}^2\right)-b_{k+1}^2-b_1^2\right).\label{overlineQ1}
    \end{align}
\end{proposition}
\par
\begin{figure}\label{Fig_star}
    \centering
    \includegraphics[scale=1.6]{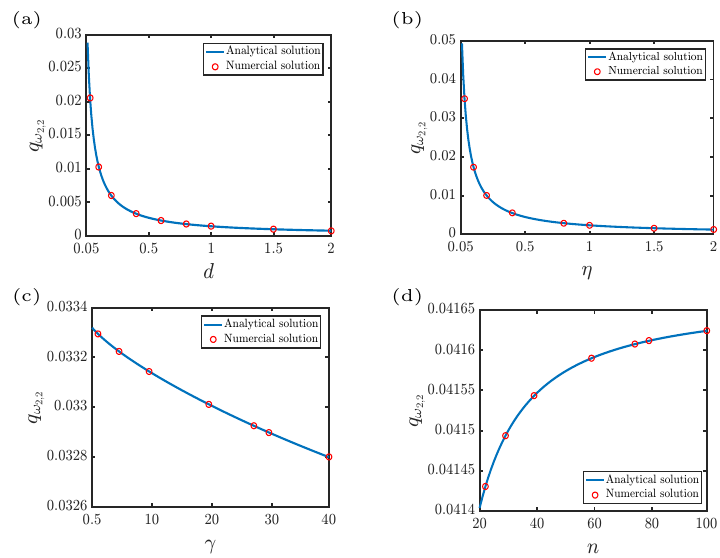}
    \caption{The dependence of the variance $q_{\omega_{2,2}}$ on the system parameters in the star graph with $b_2\neq 0$ and $b_i=0$ for $i\neq $ and indices of the nodes and lines defined in Definition \ref{definition_index}(ii).}
    \end{figure}

\begin{figure}\label{Fig_star_2}
    \centering
    \includegraphics[scale=1.6]{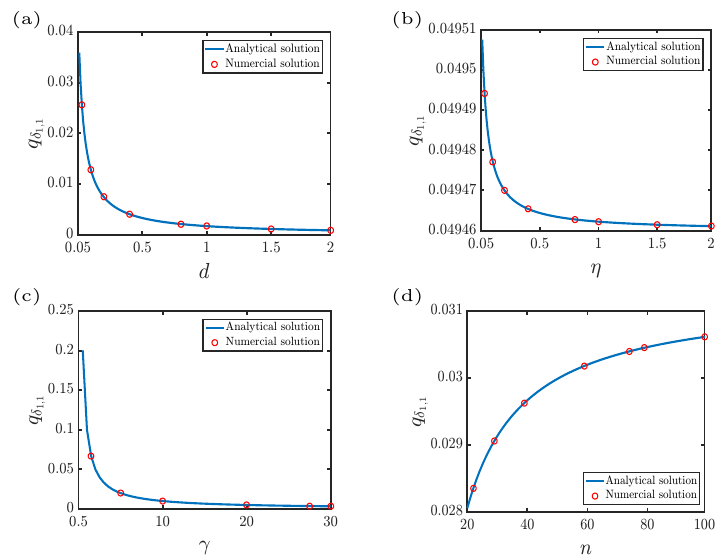}
    \caption{The dependence of variance $q_{\delta_{1,1}}$ for line $(1,2)$ on the system parameters in the star graph with $b_2\neq 0$ and $b_i=0$ for $i\neq $ and indices of the nodes and lines defined in Definition \ref{definition_index}(ii).}
    \end{figure}

\begin{figure}\label{Fig_star_3}
    \includegraphics[scale=1.6]{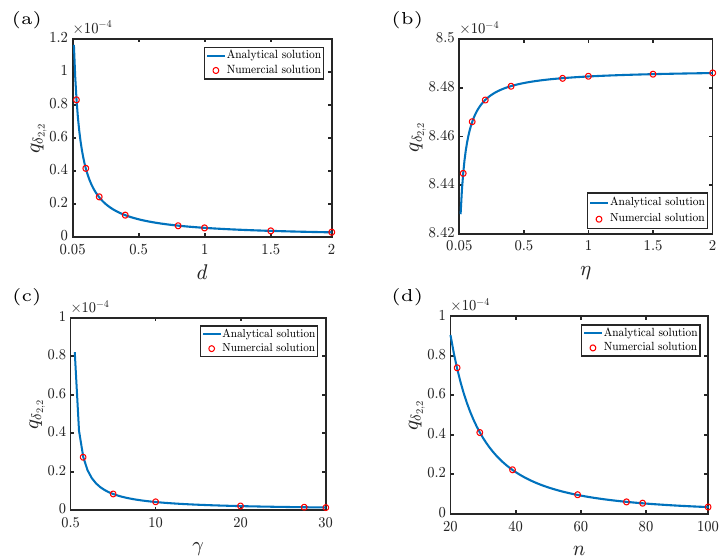}
    \caption{The dependence of variance $q_{\delta_{2,2}}$  for line $(1,3)$ on the system parameters in the star graph with $b_2\neq 0$ and $b_i=0$ for $i\neq $ and indices of the nodes and lines defined in Definition \ref{definition_index}(ii).}
    \end{figure}

 
Similar as for the complete graphs, we verify the correctness of the analytical formulas in Corollary
\ref{corollary-stargraph_2} 
by comparing the analytical solution with the numerical solution which is computed 
by Matlab from (\ref{Qy}) and (\ref{barQ}) in the star graph with $b_2\neq 0$ and $b_i=0$ for $i\neq 0$ and 
indices of the nodes and lines defined in Definition \ref{definition_index}(b). In order to satisfy Assumption 
\ref{assumption-uniformlines}, we set $P_i=0$ 
for all the nodes and $K_{ij}=K$ for all the lines, which leads to $\gamma=K\cos{\delta_{ij}^*}=K$. 
The setting of the parameters for plotting these figures are shown in Table \ref{table_star}. 
The results 
of these comparison are shown in Fig. \ref{Fig_star}, \ref{Fig_star_2} and \ref{Fig_star_3}, which demonstrates that 
these explicit formulas are all correct. 

\begin{table*}[htbp]
  \begin{center}
      \caption{The setting of the parameters for plotting Fig.\ref{Fig_star} and Fig.\ref{Fig_star_2} and Fig.\ref{Fig_star_3}.}
      \label{table_star}
      \scalebox{0.9}{
      \begin{tabular}{|c|c|c|c|c||c|c|c|c||c|c|c|c|}
      \hline
      \multirow{2}*{parameters}&\multicolumn{4}{c||}{Fig.\ref{Fig_star}}&\multicolumn{4}{c||}{Fig.\ref{Fig_star_2}}&\multicolumn{4}{c|}{Fig.\ref{Fig_star_3}}\\ \cline{2-13} 
           &(a)&(b)&(c)&(d)&(a)&(b)&(c)&(d)&(a)&(b)&(c)&(d)\\
          \hline
           $\gamma$ &10&10&$-$&10&10&10&$-$&10&10&10&$-$&10\\
          \hline
          $\eta$ &0.5&$-$&0.02&0.02&0.5&$-$&0.01&0.1&0.5&$-$&0.01&0.1\\
           \hline
          $d$ &$-$&0.3&1.2&1.5&$-$&0.2&1.2&0.4&$-$&0.2&1.2&0.4\\
           \hline
          $b_2$ &0.04&0.04&0.04&0.05&0.2&0.5&0.5&0.5&0.2&0.5&0.5&0.5\\ 
                \hline
          $n$ &20&20&30&$-$&20&10&50&$-$&20&10&50&$-$\\   
          \hline           
      \end{tabular}
      }				
  \end{center}
\end{table*}

\par 
Based on Corollary \ref{corollary-stargraph_2}, we analyze the impact of the 
system parameters on the variances of the frequency and the phase differences. 
\par 
{\bf (a) On the variance of the frequency in the star graph.}
As in 
the complete graph, the roles of the inertia $\eta$ and the damping $d$ of the synchronous machines are clear, which will not be discussed again. Here, we focus on the impacts of the weights of lines and the network size.
There are four terms in the right hand of (\ref{StarQ-omega}), i.e., the first term 
is the total amount of fluctuations caused by the disturbance at node $i=2$, which equals to 
the trace of the matrix $\mathbf{Q}_\omega$, the absolute value of the second term measures the fluctuations propagating to the root node $i=1$ and the absolute value of the 
sum of the third and the fourth term measures the fluctuations  propagating to the other $n-2$ nodes. 
\par 
First, on the influences of the weights of the lines, it yields from (\ref{StarQ-omega})
that 
\begin{align*}
\frac{\partial q_{\omega_{2,2}}}{\partial \gamma}&=-\frac{2db_2^2}{n(2d^2+\gamma\eta n)^2}-\frac{2d(n+1)(n-2)b_2^2}{n(2d^2(n+1)+\eta\gamma(n-1)^2)^2}\\
&~~~~-\frac{2d\gamma\eta(4+\gamma\eta(n+1))(n-2)b_2^2}{n(2d^2+\gamma\eta)^2(2d^2+\gamma\eta n)^2}<0,
\end{align*}
which indicates that as the weight of the lines increases, the variance of the frequency at the node with 
disturbance decrease. Thus, increasing the line capacity will accelerate 
the propagation of the fluctuations in the graph.  The lower bound of this variance is obtained as the limit as $\gamma$ increases to 
the infinity,
\begin{align*}
\lim_{\gamma\rightarrow \infty }q_{\omega_{2,2}}=\frac{b_2^2}{2d\eta}-\frac{b_2^2}{d\eta}(\frac{1}{n}-\frac{1}{n^2(n-1)^2}). 
\end{align*}
    \par 
Second, for the impact of the network size, we get from (\ref{StarQ-omega}) that for $n\geq 2$, 
\begin{align*}
\frac{\partial q_{\omega_{2,2}}}{\partial n}&=    \frac{2\gamma b_2^2(d^2+\gamma\eta n)}{d n^2(2d^2+\gamma\eta n)^2}-
    \frac{\gamma^2\eta b_2^2(4d^2+\gamma\eta n(4-n))}{dn^2(2d^2+\gamma\eta)(2d^2+\gamma\eta n)^2}\\
  +&
    \frac{2\gamma b_2^2(d^2(n^2-4n-2)+\gamma\eta(n-1)(n^2-3n+1))}{dn^2(2d^2(n+1)+\gamma\eta(n-1)^2)^2}>0,
\end{align*}
and 
\begin{align*}
\lim_{n\rightarrow \infty}q_{\omega_{2,2}}=\frac{b_2^2}{2d\eta}.
\end{align*}
The dependence of the variance $q_{\omega_{2,2}}$ on the size $n$ is shown 
in Fig.\ref{Fig_star}(d).
Because the derivative of $q_{\omega_{2,2}}$ with respect to $n$ is positive, the 
 variance of the frequency at node $i=2$ increases at the size of the network increases. 
 Note that the critical size $n_c$ in the complete graph does not exist in the star graph. 
Clearly, as the size $n$ increases to infinity, the variance $q_{\omega_{2,2}}$ converges 
to the value of the synchronous machine in the SMIB model. This shows
that for a sufficiently large size graph, the graph becomes an infinite bus connected to the 
synchronous machine. 

{\bf (b) On the variance of the phase difference in the star graph.} 
The impacts the line weight $\gamma$ and the damping coefficient $d$ on the variance 
of the phase difference, which can be observed 
from (\ref{starQ_phase1}) and (\ref{starQ_phase2}) directly,  will not be discussed here. 
We focus on the impact of the size $n$ and the inertia $\eta$. 
\par 
A new finding is that the variance also depends on the inertia in the star graph. 
By (\ref{starQ_phase1}) and (\ref{starQ_phase2}), we obtain
\begin{align*}
\frac{\partial q_{\delta_{1,1}}}{\partial \eta}=-\frac{4(n-2)db_2^2}{[2d^2(n+1)+\gamma \eta(n-1)^2]^2}\leq 0,\\
\frac{\partial q_{\delta_{k,k}}}{\partial \eta}=\frac{4db_2^2}{[2d^2(n+1)+\gamma \eta(n-1)^2]^2}> 0
\end{align*}
and 
\begin{align*}
&\lim_{\eta\rightarrow 0^+}q_{\delta_{1,1}}=\big(\frac{n-1}{2d\gamma n}-\frac{n-2}{2d\gamma n(n+1)}\big)b_2^2,~~\text{and}~~\lim_{\eta\rightarrow 0^+}q_{\delta_{k,k}}=\frac{1}{2d\gamma n(n+1)}b_2^2.
\end{align*}
From the perspective of the fluctuations of the phase angle difference, 
this demonstrates that increasing the inertia of the system, the amount of the fluctuations of the system propagating from 
the node $n=2$ to the other non-root nodes increase.  This trend can be seen in Fig.\ref{Fig_star_2}(b) and Fig. \ref{Fig_star_3}(b). 
This is different from the findings in the network with uniform damping-disturbance ratio, where 
the inertia have no impact on the variances of the phase angle differences \cite{ZhenWangAutom}. 
\par 
Regarding to the influence of the network size $n$, we derive
\begin{align*}
   \frac{\partial q_{\delta_{1,1}}}{\partial n}
    &=
    \frac{4d^4(2n^2-2n-1)+4d^2\gamma\eta(2n^3-6n^2+n-1)}{2d\gamma n^2(2d^2(1+n)+\gamma\eta(n-1)^2)^2}\\
    &~~~~~~+\frac{\gamma^2\eta^2(n-1)(2n^3-4n^2-3n+1)}
    {2d\gamma n^2(2d^2(1+n)+\gamma\eta(n-1)^2)^2}b_2^2>0,\\
        \frac{\partial q_{\delta_{kk}}}{\partial n}
    &=
    -\frac{4d^4(1+2n)+4d^2\gamma\eta(2n^2-n+1)+\gamma^2\eta^2(n-1)(2n^2+3n-1)}
    {2d\gamma n^2(2d^2(1+n)+\gamma\eta(n-1)^2)^2}b_2^2<0.
\end{align*}
This indicates that in the star graph the fluctuations the line connecting to the 
source node of the disturbance will increase as the graph size increases, while 
the fluctuations in other lines will decreases. 
These trends can also be observed in Fig. \ref{Fig_star_2}(d) and Fig.\ref{Fig_star_3}(d). 
\par 
Comparing the formulas of $\mathbf{Q}_\delta$ in Proposition \ref{Corollary-star} and that of $\overline{\mathbf{Q}}_\delta$ in Proposition \ref{proposition_kuramoto}, it is found that 
\begin{align*}
\lim_{\eta\rightarrow 0}\mathbf{Q}_\delta=\overline{\mathbf{Q}}_\delta, 
\end{align*}
where $\eta\rightarrow 0$ means that the inertia goes to zero. This property demonstrates that the variance of the phase difference in a power system with very small inertia 
can be estimated by that in the non-uniform Kuramoto model of a star graph. 
\par 

\subsection{Summation of the findings in the complete graph and the star graph}
Since the formulas of the variance matrices in Corollary \ref{corollay-complete_graph} and \ref{corollary-stargraph-1} are the same, then the trend of the propagation of the disturbance from the root node 
to the other nodes in a star graph is the same as the propagation of the disturbance from 
a node to others in the complete graph which has the size of the star graph. This will not be
discussed again. 
In this subsection, we summarize the findings from Corollary \ref{corollay-complete_graph} for the complete graphs and 
Corollary \ref{corollary-stargraph_2} for the star graph. 
\par 
First, it has been found from these corollaries that increasing the 
line capacity will accelerate the propagation of the fluctuations in the network.
Second, as the size of the graph increases, the graph and the source 
node of the disturbance play their roles as an infinite bus and a single machine respectively in a SMIB model.  
Third, the impacts of the inertia of the synchronous machine on the 
phase differences depends on the network topology. It has no impact on the variance of the phase difference in the complete 
graph while it may impact the propagation of the fluctuations in the phase 
difference from the source node of the disturbance to the other nodes. 

\section{The Proofs}\label{Section:proofs}
In (\ref{blockmatrix}), $\mathbf{A}_2$ and $\mathbf{B}_2$ are further decomposed as,
\begin{align}\label{matrix-A2}
\mathbf{A}_2=
\begin{bmatrix}
\mathbf{0} &\mathbf{A}_{22}\\
\mathbf{A}_{23} &\mathbf{A}_{24}
\end{bmatrix},~~
\mathbf{B}_2=
\begin{bmatrix}
\mathbf{0}\\
\mathbf{B}_{22}
\end{bmatrix},
\end{align}
where
\begin{subequations}\label{matrix-A22}
\begin{align}
&\mathbf{A}_{22}=
\begin{bmatrix}
\mathbf{0}&\mathbf{I}_{n-1}
\end{bmatrix}
\in\mathbb{R}^{(n-1)\times n}, ~~
\mathbf{A}_{23}^{\bm\top}=
\begin{bmatrix}
\mathbf{0}&-\mathbf{\Lambda}_{n-1}
\end{bmatrix}
\in\mathbb{R}^{(n-1)\times n},\\
&\mathbf{A}_{24}=-\mathbf{U}^{\top}\mathbf{M}^{-1}\mathbf{D}\mathbf{U}\in\mathbb{R}^{n\times n},~~ \mathbf{B}_{22}=\mathbf{U}^{\top}\mathbf{M}^{-1/2} \widetilde{\mathbf{B}}\in\mathbb{R}^{n\times n}.
\end{align}
\end{subequations}
Here, $\bm \Lambda_{n-1}=\text{diag}(\lambda_i,i=2,\cdots,n)\in\mathbb{R}^{(n-1)\times(n-1)}$ is obtained by removing the first column and the first row of the diagonal matrix $\bm \Lambda_n$.

\emph{Proof of Theorem \ref{maintheorem_kuramot}}
\par 
With the matrix $\mathbf{C}_2$ in (\ref{C2-delta-omega}),  we obtain from (\ref{Qy}) that
\begin{equation*}
 \begin{aligned}
            \mathbf{Q}_y
             =\mathbf{C}_{2}\mathbf{Q}_{x}\mathbf{C}_{2}^{\bm \top}=
            \begin{bmatrix}
                \widetilde{\mathbf{C}}^\top\mathbf{M}^{-1/2}\widehat{\mathbf{U}}\mathbf{G}\widehat{\mathbf{U}}^{\bm \top}\mathbf{M}^{-1/2}\widetilde{\mathbf{C}}
                &~~
                \widetilde{\mathbf{C}}^{\bm \top}\mathbf{M}^{-1/2}\widehat{\mathbf{U}}\mathbf{S}\mathbf{U}^{\bm \top}\mathbf{M}^{-1/2} 
                \\
                \mathbf{M}^{-1/2}\mathbf{U}\mathbf{S}^{\bm\top}\widehat{\mathbf{U}}^{\bm \top}\mathbf{M}^{-1/2}\widetilde{\mathbf{C}}
                &
                \mathbf{M}^{-1/2}\mathbf{U}\mathbf{R}\mathbf{U}^{\bm\top}\mathbf{M}^{-1/2}
            \end{bmatrix}
        \end{aligned}
    \end{equation*}
With the block matrices $\mathbf{A}_2$ and $\mathbf{B}_2$ in (\ref{matrix-A2}) and the blocks $\mathbf{A}_{22}$, $\mathbf{A}_{23}$, $\mathbf{A}_{24}$ and $\mathbf{B}_{22}$ in (\ref{matrix-A22}) and the block matrix $\mathbf{Q}_x$ in (\ref{equationQx}), we derive 
from the Lyapunov equation (\ref{barQ}) that 
\begin{align*}
\begin{bmatrix}
\mathbf{0}&\mathbf{A}_{22}\\
\mathbf{A}_{23}&\mathbf{A}_{24}
\end{bmatrix}
\begin{bmatrix}
\mathbf{G}&\mathbf{S}\\
\mathbf{S}^{\bm\top}&\mathbf{R}
\end{bmatrix}
&+
\begin{bmatrix}
\mathbf{G}&\mathbf{S}\\
\mathbf{S}^{\bm\top}&\mathbf{R}
\end{bmatrix}
\begin{bmatrix}
\mathbf{0}&\mathbf{A}_{22}\\
\mathbf{A}_{23}&\mathbf{A}_{24}
\end{bmatrix}^{\bm\top}+
\begin{bmatrix}
\mathbf{0}\\
\mathbf{B}_{22}
\end{bmatrix}
\begin{bmatrix}
\mathbf{0}&\mathbf{B}_{22}^{\bm\top}
\end{bmatrix}
=\mathbf{0}
\end{align*}
which yields
\begin{subequations}\label{equivalenteq}
\begin{align}
\mathbf{S}\mathbf{A}_{22}^{\top}+\mathbf{A}_{22}\mathbf{S}^{\top}&=\mathbf{0},\label{equivalent1}\\
\mathbf{G}\mathbf{A}_{23}^{\top}+\mathbf{S}\mathbf{A}_{24}^{\top}+\mathbf{A}_{22}\mathbf{R}&=\mathbf{0},\label{equivalent2}\\
\mathbf{S}^{\top}\mathbf{A}_{23}^{\top}+\mathbf{R}\mathbf{A}_{24}^{\top}+\mathbf{A}_{23}\mathbf{S}+\mathbf{A}_{24}\mathbf{R}&=-\mathbf{B}_{22}\mathbf{B}_{22}^{\top}.\label{equivalent3}
\end{align}
\end{subequations}
Denote $\mathbf{S}=
        \begin{bmatrix}
            \mathbf{S}_{1} & \mathbf{S}_{2}
        \end{bmatrix}$ with $\mathbf{S}_{1}\in\mathbb{R}^{n-1}$ and $\mathbf{S}_{2}\in\mathbb{R}^{(n-1)\times(n-1)}$ and 
insert it into (\ref{equivalent1}), then
 \begin{equation}
        \begin{bmatrix}
            \mathbf{S}_{1} & \mathbf{S}_{2}
        \end{bmatrix}
        \begin{bmatrix}
            \mathbf{0}\\
            \mathbf{I}_{n-1}
        \end{bmatrix}
        +
        \begin{bmatrix}
            \mathbf{0} & \mathbf{I}_{n-1}
        \end{bmatrix}
        \begin{bmatrix}
            \mathbf{S}_{1}^{ \top}\\
            \mathbf{S}_{2}^{ \top}
        \end{bmatrix}
        =\mathbf{0}.
    \end{equation}
which leads to $$\mathbf{S}_{2}+\mathbf{S}_{2}^{\top}=\mathbf{0},$$
which means that $\mathbf{S}_{2}$ is a skew-symmetric matrix. Thus, the elements of $\mathbf{S}$ satisfy
\begin{align*}
s_{j-1,i+1}=-s_{i,j},i=1,2,\cdots,n-1,~j=2,\cdots,n
\end{align*}
It yields from Assumption \ref{assumption} and (\ref{matrix-A22}) that
$\mathbf{A}_{24}=-\alpha\mathbf{I}_{n}$. 
Hence, we obtain from  (\ref{equivalent2}) and (\ref{equivalent3}) that 
\begin{subequations}
        \begin{align}
            \alpha\mathbf{S}&=\mathbf{G}\mathbf{A}_{23}^{\top}+\mathbf{A}_{22}\mathbf{R},
            \label{eq:b1}\\
            2\alpha\mathbf R&=\mathbf{S}^{ \top}\mathbf{A}_{23}^{\top}+\mathbf{A}_{23}\mathbf{S}+\mathbf{B}_{22}\mathbf{B}_{22}^{ \top}
            \label{eq:c1}.
        \end{align}
    \end{subequations}
By inserting (\ref{eq:c1}) into (\ref{eq:b1}), we derive
\begin{align*}
2\alpha^{2}\mathbf{S}&=2\alpha\mathbf{G}\mathbf{A}_{23}^{\top}+\mathbf{A}_{22}\mathbf{S}^{ \top}\mathbf{A}_{23}^{ \top}+\mathbf{A}_{22}\mathbf{A}_{23}\mathbf{S}+\mathbf{A}_{22}\mathbf{B}_{22}\mathbf{B}_{22}^{ \top}\\
&~~~~\text{by (\ref{equivalent1})}\\ 
&=2\alpha\mathbf{G} \mathbf{A}_{23}^{\top}-\mathbf{S}\mathbf{A}_{22}^{\top}
        \mathbf{A}_{23}^{\top}+\mathbf{A}_{22}\mathbf{A}_{23}\mathbf{S}+\mathbf{A}_{22}\mathbf{B}_{22}\mathbf{B}_{22}^{\bm \top}.
\end{align*}
Plugging $\mathbf{A}_{23}$ and $\mathbf{A}_{22}$ of (\ref{matrix-A22}) into the above equation, we get 
 \begin{equation*}
        \begin{aligned}
            &2\alpha\mathbf{G}
            \begin{bmatrix}
                \mathbf{0} & -\mathbf{\Lambda}_{n-1}
            \end{bmatrix}
            +\begin{bmatrix}
                \mathbf{0} & \mathbf{I}_{n-1}
            \end{bmatrix}
            \mathbf{B}_{22}\mathbf{B}_{22}^{\top}
            =2\alpha^{2}\mathbf{S}+\mathbf{S}
            \begin{bmatrix}
                \mathbf{0} & \mathbf{0}\\
                \mathbf{0} & -\mathbf{\Lambda}_{n-1}
            \end{bmatrix}
            +\mathbf{\Lambda}_{n-1}\mathbf{S}.
        \end{aligned}
    \end{equation*}
With the notation of $\mathbf{S}=
        \begin{bmatrix}
            \mathbf{S}_{1} & \mathbf{S}_{2}
        \end{bmatrix}$, we obtain from the above equation that  
\begin{equation}\label{eq:block}
        \begin{aligned}
            \begin{bmatrix}
                \mathbf{0} & -2\alpha\mathbf{G}\mathbf{\Lambda}_{n-1}
            \end{bmatrix}
            +&\begin{bmatrix}
                \mathbf{0} & \mathbf{I}_{n-1}
            \end{bmatrix}
            \mathbf{B}_{22}\mathbf{B}_{22}^{\top}\\
            &=
            2\alpha^2
            \begin{bmatrix}
               \mathbf{S}_{1} & \mathbf{S}_{2} 
            \end{bmatrix}
            +
            \begin{bmatrix}
                \mathbf{\Lambda}_{n-1}\mathbf{S}_{1} & \mathbf{\Lambda}_{n-1}\mathbf{S}_{2}
            \end{bmatrix}
            +
            \begin{bmatrix}
                \mathbf{0} & -\mathbf{S}_{2}\mathbf{\Lambda}_{n-1}
            \end{bmatrix}\\
            &=
            \begin{bmatrix}
                2\alpha^2\mathbf{S}_{1}+\mathbf{\Lambda}_{n-1}\mathbf{S}_{1} & 2\alpha^2\mathbf{S}_{2}+\mathbf{\Lambda}_{n-1}\mathbf{S}_{2}-\mathbf{S}_{2}\mathbf{\Lambda}_{n-1} 
            \end{bmatrix}.
        \end{aligned}
    \end{equation}
From the definition of $\mathbf{B}_{22}$ in (\ref{matrix-A22}),  we obtain
    \begin{equation}\label{matrixB22timesB22}
        \begin{aligned}
\begin{bmatrix}
                \mathbf{0} & \mathbf{I}_{n-1}
            \end{bmatrix}
            \mathbf{B}_{22}\mathbf{B}_{22}^{\top}
            &=
            \begin{bmatrix}
                \sum\limits_{k}u_{k,2}u_{k,1}\xi_{k} & \sum\limits_{k}u_{k,2}^2\xi_{k} & 
                \cdots & \sum\limits_{k}u_{k,2}u_{k,n}\xi_{k}\\
                \sum\limits_{k}u_{k,3}u_{k,1}\xi_{k} & \sum\limits_{k}u_{k,3}u_{k,2}\xi_{k} & 
                \cdots & \sum\limits_{k}u_{k,3}u_{k,n}\xi_{k}\\
                \vdots & \vdots & \vdots & \vdots\\
                \sum\limits_{k}u_{k,n}u_{k,1}\xi_{k} & \sum\limits_{k}u_{k,n}u_{k,2}\xi_{k} & 
                \cdots & \sum\limits_{k}u_{k,n}^2\xi_{k}
            \end{bmatrix}
        \end{aligned}
    \end{equation}
    where  $u_{i,j}$ is the element of the matrix $\mathbf{U}$ and $\xi_{k}$ represent the $k$-th diagonal elements in $\mathbf{M}^{-1/2}\widetilde{\mathbf{B}}^2\mathbf{M}^{-1/2}$. 
    Plugging (\ref{matrixB22timesB22}) into (\ref{eq:block}),  we obtain that 
    the elements of the vector $2\alpha^2\mathbf{S}_{1}+\mathbf{\Lambda}_{n-1}\mathbf{S}_{1} $ satisfy
 \begin{equation*}
        \begin{aligned}
            \begin{bmatrix}
                (2\alpha^2+\lambda_2)s_{{1,1}}\\
                (2\alpha^2+\lambda_3)s_{{2,1}}\\
                \vdots\\
                (2\alpha^2+\lambda_n)s_{{n-1,1}}
            \end{bmatrix}
            &=
            \begin{bmatrix}
                \sum_{k}u_{k,2}u_{k,1}\xi_{k} \\
                \sum_{k}u_{k,3}u_{k,1}\xi_{k} \\
                \vdots\\
                \sum_{k}u_{k,n}u_{k,1}\xi_{k}
            \end{bmatrix}.
        \end{aligned}                 
    \end{equation*}
which yields (\ref{eq:result1}).
Similarly,  the elements of the matrix $2\alpha^2\mathbf{S}_{2}+\mathbf{\Lambda}_{n-1}\mathbf{S}_{2}-\mathbf{S}_{2}\mathbf{\Lambda}_{n-1}$ satisfy 
 \begin{equation}
        \label{eq:Q1}
        \begin{aligned}
            &\begin{bmatrix}
                0 & (-2\alpha^2\!-\!\lambda_2+\lambda_3)s_{{2,2}} &\cdots & (-2\alpha^2-\lambda_2+\lambda_n)s_{{n-1,2}}\\
                (2\alpha^2+\lambda_3-\lambda_2)s_{{2,2}} & 0 &\cdots &(-2\alpha^2-\lambda_3+\lambda_n) s_{{n-1,3}}\\
                \vdots & \vdots & \vdots &\vdots \\
                (2\alpha^2+\lambda_n-\lambda_2)s_{{n-1,2}} & (2\alpha^2+\lambda_n-\lambda_3)s_{{n-1,3}} & \cdots & 0
            \end{bmatrix}\\
            =
            &\begin{bmatrix}
                \sum\limits_{k}u_{k,2}^2\xi_{k} & \sum\limits_{k}u_{k,2}u_{k,3}\xi_{k} & \cdots & \sum\limits_{k}u_{k,2}u_{k,n}\xi_{k}\\
                \sum\limits_{k}u_{k,3}u_{k,2}\xi_{k} & \sum\limits_{k}u_{k,3}^2\xi_{k} & \cdots & \sum\limits_{k}u_{k,3}u_{k,n}\xi_{k}\\
                \vdots & \vdots & \vdots & \vdots \\
                \sum\limits_{k}u_{k,n}u_{k,2}\xi_{k} & \sum\limits_{k}u_{k,n}u_{k,3}\xi_{k} & \cdots & \sum\limits_{k}u_{k,n}^2\xi_{k}
            \end{bmatrix}\\
            &-2\alpha
            \begin{bmatrix}
                \lambda_{2}g_{{1,1}} & \lambda_{3}g_{{1,2}} & \cdots & \lambda_{n}g_{{1,n-1}}\\
                \lambda_{2}g_{{2,1}} & \lambda_{3}g_{{2,2}} & \cdots & \lambda_{n}g_{{2,n-1}}\\
                \vdots & \vdots & \vdots & \vdots\\
                \lambda_{2}g_{{n-1,1}} & \lambda_{3}g_{{n-1,2}} & \cdots & \lambda_{n}g_{{n-1,n-1}}\\
            \end{bmatrix}
        \end{aligned}
    \end{equation}
By the symmetry of $\mathbf{G}$, i.e., $g_{{i,j}}=g_{{j,i}}$, we obtain from (\ref{eq:Q1}) that for $i=1,2,\cdots,n-1,j=2,\cdots,n$, 
    \begin{equation}
        \label{eq:Q22}
        \begin{aligned}
            &\left(
                2-\frac{2\alpha^2}{\lambda_{i+1}}-\frac{2\alpha^2}{\lambda_{j}}-\frac{\lambda_{i+1}}{\lambda_{j}}-\frac{\lambda_{j}}{\lambda_{i+1}}
             \right)s_{{i,j}}
            =\left(
                \frac{1}{\lambda_{i+1}}-\frac{1}{\lambda_{j}}
              \right)\mathbf{u}_{i+1}^{ \top}\mathbf{M}^{-1/2}\widetilde{\mathbf{B}}^2\mathbf{M}^{-1/2}\mathbf{u}_{j}. 
        \end{aligned}
    \end{equation}
which yields (\ref{eq:result2}). 
  \par 
  From (\ref{eq:Q1}),
   we obtain for $i=1,2,\cdots,n,$
       \begin{equation}
        \label{eq:Q11}
        g_{{i,i}}=\frac{1}{2\alpha\lambda_{i+1}}\mathbf{u}_{i+1}^{\top}\mathbf{M}^{-1/2}\widetilde{\mathbf{B}}^2\mathbf{M}^{-1/2}\mathbf{u}_{i+1},
        \, i=1,2,\cdots,n-1.
    \end{equation}
   and for 
    $ i=1,2,\cdots,n-1, j=i+1,\cdots,n-1$, 
   \begin{equation}
        \label{eq:Q12}
        \begin{aligned}
            -2\alpha\lambda_{j+1}g_{{i,j}}
            =(\lambda_{j+1}-\!\lambda_{i+1}\!-2\alpha^2)s_{{j,i+1}}-\mathbf{u}_{i+1}^{\top}\mathbf{M}^{-1/2}\widetilde{\mathbf{B}}^2\mathbf{M}^{-1/2}\mathbf{u}_{j+1},
        \end{aligned}
    \end{equation}
    which yield (\ref{eq:result3}) with the expression of $\mathbf{S}$ in (\ref{eq:result2}).
 \par 
Now, we focus on the derivation of $\mathbf{R}$. We denote 
 \begin{equation*}
        \mathbf{R}=
        \begin{bmatrix}
            R_{1} & \mathbf{R}_{2}^{\top}\\
            \mathbf{R}_{2} & \mathbf{R}_{3}
        \end{bmatrix}
    \end{equation*}
     where $R_{1}\in\mathbb{R}$, $\mathbf{R}_{2}\in\mathbb{R}^{(n-1)}$ and $\mathbf{R}_{3}\in\mathbb{R}^{(n-1)\times (n-1)}$. Then, 
(\ref{eq:c1}) is rewritten into
 \begin{equation}\label{eq:Q3last}
        \begin{aligned}
            \begin{bmatrix}
                R_{1} & \mathbf{R}_{2}^{\top}\\
                \mathbf{R}_{2} & \mathbf{R}_{3}
            \end{bmatrix}
            =\displaystyle{\frac{1}{2\alpha}}\left(
                \begin{bmatrix}
                    0 & -\mathbf{S}_{1}^{\top}\mathbf{\Lambda}_{n-1}\\
                    -\mathbf{\Lambda}_{n-1}\mathbf{S}_{1} &~-\mathbf{\Lambda}_{n-1}\mathbf{S}_{2}-\mathbf{S}_{2}^{ \top}\mathbf{\Lambda}_{n-1}
                \end{bmatrix} 
                +\mathbf{B}_{22}\mathbf{B}_{22}^{\top}
            \right).
        \end{aligned}
    \end{equation}
 where 
    \begin{equation*}
       \mathbf{B}_{22}\mathbf{B}_{22}^{\top}=
        \begin{bmatrix}
            \sum\limits_{k}u_{k,1}^2\xi_{k} & \sum\limits_{k}u_{k,1}u_{k,2}\xi_{k} & 
            \cdots & \sum\limits_{k}u_{k,1}u_{k,n}\xi_{k}\\
            \sum\limits_{k}u_{k,2}u_{k,1}\xi_{k} & \sum\limits_{k}u_{k,2}^2\xi_{k} & 
            \cdots & \sum\limits_{k}u_{k,2}u_{k,n}\xi_{k}\\
            \vdots & \vdots & \vdots & \vdots\\
            \sum\limits_{k}u_{k,n}u_{k,1}\xi_{k} & \sum\limits_{k}u_{k,n}u_{k,2}\xi_{k} & 
            \cdots & \sum\limits_{k}u_{k,n}^2\xi_{k}
        \end{bmatrix}.
    \end{equation*} 
 From (\ref{eq:Q3last}), we obtain the expression of $R_1$ which equals 
 to $r_{1,1}$ in (\ref{eq:result4}). 
From (\ref{eq:Q3last}), we obtain, 
    \begin{equation*}
        \begin{aligned}
            \mathbf{R}_{2}
            =\displaystyle{\frac{1}{2\alpha}}\left(-
            \begin{bmatrix}
                \lambda_{2}s_{{1,1}} \\
                \lambda_{3}s_{{2,1}} \\
                \vdots \\
                \lambda_{n}s_{{n-1,1}}
            \end{bmatrix}
            +
            \begin{bmatrix}
                \sum_{k}u_{k,1}u_{k,2}\xi_{k} \\
                \sum_{k}u_{k,1}u_{k,3}\xi_{k} \\
                \vdots \\
                \sum_{k}u_{k,1}u_{k,n}\xi_{k}
            \end{bmatrix}
            \right).
        \end{aligned}
    \end{equation*}
  from which we obtain for $i=\!1,2,\cdots,n-1$,
      \begin{equation}
        \label{eq:Q33}
        r_{{i+1,1}}\!=\frac{1}{2\alpha}(-\lambda_{i+1}s_{{i,1}}+\mathbf{u}_{i+1}^{\top}\bm M^{-1/2}\widetilde{\mathbf{B}}^2\mathbf{M}^{-1/2}\mathbf{u}_{1}),
    \end{equation}
    which leads (\ref{eq:result5}) with the expression of $s_{i,1}$ in (\ref{eq:result1}). 
 \par 
From (\ref{eq:Q3last}), we further obtain, 
    \begin{equation}
        \begin{aligned}
           2\alpha \mathbf{R}_{3}
            =
                &
                \begin{bmatrix}
                    0 & (\lambda_{2}-\lambda_3)s_{{2,2}}  & \cdots & (\lambda_{2}-\lambda_n)s_{{n-1,2}}\\
                    (-\lambda_{3}+\lambda_2)s_{{2,2}} & 0 & \cdots & (\lambda_{3}-\lambda_n) s_{{n-1,3}}\\
                    \vdots & \vdots & \vdots &\vdots & \vdots \\
                    (-\lambda_{n}+\lambda_2)s_{{n-1,2}} & (-\lambda_{n}+\lambda_3)s_{{n-1,3}} & \cdots & 0
                \end{bmatrix} \\
                &  +
                \begin{bmatrix}
                    \sum\limits_{k}u_{k,2}^2\xi_{k} & \sum\limits_{k}u_{k,2}u_{k,3}\xi_{k} & 
                    \cdots & \sum\limits_{k}u_{k,2}u_{k,n}\xi_{k}\\
                    \sum\limits_{k}u_{k,3}u_{k,2}\xi_{k} & \sum\limits_{k}u_{k,3}^2\xi_{k} & 
                    \cdots & \sum\limits_{k}u_{k,3}u_{k,n}\xi_{k}\\
                    \vdots & \vdots & \vdots & \vdots\\
                    \sum\limits_{k}u_{k,n}u_{k,2}\xi_{k} & \sum\limits_{k}u_{k,n}u_{k,3}\xi_{k} & 
                    \cdots & \sum\limits_{k}u_{k,n}^2\xi_{k}
                \end{bmatrix}
        \end{aligned}
    \end{equation}
Thus,  for $i,j=2,3,\cdots,n$,
    \begin{equation*}
        \begin{aligned}
            r_{{i,j}}&=\frac{1}{2\alpha}\left(
                (\lambda_{i}-\lambda_{j})s{_{j-1,i}}+\mathbf{u}_{i}^{\bm \top}\mathbf{M}^{-1/2}\widetilde{\mathbf{B}}^2\mathbf{M}^{-1/2}\mathbf{u}_{j}
                            \right)
        \end{aligned}
    \end{equation*}
    which leads to (\ref{eq:result5}) with the formula of $s_{j-1,i}$ in (\ref{eq:result2}). \hfill $\square$
    \par 
    
\emph{Proof of Proposition \ref{Corollary-complete}}.
\par 
 Following Lemma \ref{lemma_eigenvalues} and Assumption \ref{assumption2} and \ref{assumption-uniformlines}, we obtain the eigenvalues of 
the matrix $\mathbf{M}^{-1/2}\mathbf{L}\mathbf{M}^{-1/2}$ as defined in (\ref{decomposition}), which satisfy 
 \begin{align*}
 \lambda_1=0,\lambda_i=\gamma\eta^{-1} n,~\text{for}~i=2,\cdots,n. 
 \end{align*}
 With these eigenvalues and Theorem \ref{maintheorem_kuramot}, the formula 
 of $\mathbf{R}$ is rewritten into
    \begin{align}\label{eigenvalue_complete}
        \mathbf{R}=\eta^{-1}
        \begin{bmatrix}
            \frac{1}{2\alpha}\mathbf{u}_1^\top \widetilde{\mathbf{B}}^2\mathbf{u}_1 & \frac{\alpha}{2\alpha^2+\eta^{-1}\gamma n}\mathbf{u}_1^\top\widetilde{\mathbf{B}}^2\widehat{\mathbf{U}}\\
            \frac{\alpha}{2\alpha^2+\eta^{-1}\gamma n}\widehat{\mathbf{U}}^\top\widetilde{\mathbf{B}}\mathbf{u}_1 & \frac{1}{2\alpha}\widehat{\mathbf{U}}^\top \widetilde{\mathbf{B}}^2\widehat{\mathbf{U}} 
        \end{bmatrix}. 
    \end{align}
    Hence,  the variance matrix of frequency satisfies
    \begin{align*}
        \mathbf{Q}_{\bm\omega}&=\eta^{-2}
        \begin{bmatrix}
            \mathbf{u}_1 & \widehat{\mathbf{U}}
        \end{bmatrix}
        \begin{bmatrix}
            \frac{1}{2\alpha}\mathbf{u}_1^\top \widetilde{\mathbf{B}}^2\mathbf{u}_1 & \frac{\alpha}{2\alpha^2+\eta^{-1}\gamma n}\mathbf{u}_1^\top\widetilde{\mathbf{B}}^2\widehat{\mathbf{U}}\\
            \frac{\alpha}{2\alpha^2+\eta^{-1}\gamma n}\widehat{\mathbf{U}}^\top\widetilde{\mathbf{B}}^2\mathbf{u}_1 & \frac{1}{2\alpha}\widehat{\mathbf{U}}^\top \widetilde{\mathbf{B}}^2\widehat{\mathbf{U}} 
        \end{bmatrix}
        \begin{bmatrix}
            \mathbf{u}_1 & \widehat{\mathbf{U}}
        \end{bmatrix}^\top \\
       & \text{by}~~ \mathbf{u}_1\mathbf{u}_1^\top+\widehat{\mathbf{U}}\widehat{\mathbf{U}}^\top=\mathbf{I}\\
       & =\eta^{-2}
        \left(
            \frac{1}{2\alpha}\widetilde{\mathbf{B}}^2+
            \left(
                \frac{\alpha}{2\alpha^2+\eta^{-1}\gamma n}-\frac{1}{2\alpha}
            \right)
            \left(
                \mathbf{u}_1\mathbf{u}_1^\top\widetilde{\mathbf{B}}^2+
                \widetilde{\mathbf{B}}^2\mathbf{u}_1\mathbf{u}_1^\top
            \right)
        \right.\\
            &+
            \left.
            \left(
                \frac{1}{\alpha}-\frac{2\alpha}{2\alpha^2+\eta^{-1}\gamma n}
            \right)\mathbf{u}_1\mathbf{u}_1^\top\widetilde{\mathbf{B}}^2\mathbf{u}_1\mathbf{u}_1^\top
        \right).
    \end{align*}
Inserting $\mathbf{u}_1=1/\sqrt{n}\mathbf{1}_n^\top$ into the above equation, we obtain the diagonal elements
of $\mathbf{Q}_{\omega}$, which satisfy for $i=1,\cdots,n$, 
    \begin{align*}
        q_{\omega_{i,i}}&=\eta^{-2}
        \left(
            \frac{b_i^2}{2\alpha}+
            \frac{-\gamma b_i^2}{d(2\alpha^2+\eta^{-1}\gamma n)}
            +\frac{\gamma\text{tr}\left(\widetilde{\mathbf{B}}^2\right)}{dn(2\alpha^2+\eta^{-1}\gamma n)}
        \right)\\
        &=\big[\frac{1}{2d\eta}-\frac{\gamma(n-1)}{dn(2d^2+\gamma\eta n)}\big]b_i^2
        +\frac{\gamma}{d n\left(2d^2+\gamma \eta n\right)}(\text{tr}(\widetilde{\mathbf{B}}^2)-b_{i}^2).
    \end{align*}    
With the eigenvalues in (\ref{eigenvalue_complete}) and 
 the formula of $\mathbf{G}$ in (\ref{eq:result3}), we derive
    \begin{equation*}
        \begin{aligned}
            \mathbf{G}=\frac{1}{2\alpha\gamma n}\mathbf{\widehat{U}}^\mathsf{T}\widetilde{\mathbf{B}}^2 {\widehat{\mathbf{U}}}.
        \end{aligned}
        \end{equation*}  
 which is inserted into (\ref{eq:variance0}),  we obtain
\begin{equation}
\begin{aligned}
\mathbf{Q}_{\delta}=&
\frac{1}{2\alpha\eta\gamma n}\widetilde{\mathbf{C}}^\top\widehat{\mathbf{U}}\widehat{\mathbf{U}}^\top\widetilde{\mathbf{B}}^2\widehat{\mathbf{U}}\widehat{\mathbf{U}}^\top \widetilde{\mathbf{C}}\\
&\text{by} [\mathbf{u}_1~~\widehat{\mathbf{U}}][\mathbf{u}_1~~\widehat{\mathbf{U}}]^\top=\mathbf{\mathbf{u}}_1\mathbf{\mathbf{u}}_1^\top+\widehat{\mathbf{U}}\widehat{\mathbf{U}}^\top=\mathbf{I}_n\\
=&\frac{1}{2d\gamma n}\widetilde{\mathbf{C}}^\top(\mathbf{I}_n-\mathbf{u}_1\mathbf{u}_1^\top)\widetilde{\mathbf{B}}^2(\mathbf{I}_n-\mathbf{u}_1\mathbf{u}_1^\top)\widetilde{\mathbf{C}}\\
&\text{by}~~\widetilde{\mathbf{C}}^\top\mathbf{u}_1=\mathbf{0}\\
=&\frac{1}{2d\gamma n}\widetilde{\mathbf{C}}^\top \widetilde{\mathbf{B}}^2\widetilde{\mathbf{C}}
\end{aligned}
\end{equation} 
which leads to (\ref{completenetowrk2}).    
If we substitute the formula of incidence matrix into the above equation, we will get (\ref{eq:Q_delta}). \hfill $\square$\par  
\emph{Proof of Proposition \ref{proposition47}}
\par 
Following Lemma \ref{lemma_eigenvalues}(i) and the assumption of $\mathbf{D}=d\mathbf{I}$
and the weight $K_{i,j}\cos\delta_{ij}^*=\gamma$ for all the lines, we obtain the eigenvalues of 
the matrix $\mathbf{D}^{-1/2}\mathbf{L}\mathbf{D}^{-1/2}$, 
\begin{align*}
\overline{\lambda}_1=0,~~\overline{\lambda}_i=n\gamma/d~\text{for}~i=2,\cdots,n. 
\end{align*}
Plugging these eigenvalues of the Laplacian matrix of the complete graph into (\ref{Q1}), we obtain
the expression of the elements of the matrix $\overline{\mathbf{Q}}_{x}$, 
$$\overline{q}_{x_{ij}}=\frac{1}{2\gamma n}\overline{\mathbf{u}}_{i+1}^{\top}\widetilde{\mathbf{B}}^2\overline{\mathbf{u}}_{j+1},\, \forall i,\, j=1,\cdots,n-1$$
Thus,  $\overline{\mathbf{Q}}_{x}=\frac{1}{2\gamma n}\overline{\mathbf{U}}_2^\top\widetilde{\mathbf{B}}^2\overline{\mathbf{U}}_2$. Following (\ref{Qdelta}), we derive
    \begin{align*}
        \overline{\mathbf{Q}}_{y}&=\frac{1}{2d\gamma n}\widetilde{\mathbf{C}}^{\top}\overline{\mathbf{U}}_2\overline{\mathbf{U}}_2^\top\widetilde{\mathbf{B}}^2\overline{\mathbf{U}}_2\overline{\mathbf{U}}_2^{\top}\widetilde{\mathbf{C}}
        =\frac{1}{2d\gamma n}\widetilde{\mathbf{C}}^{\top}(\mathbf I-\overline{\mathbf{u}}_{1}\overline{\mathbf{u}}_{1}^\top)\widetilde{\mathbf{B}}^2(\mathbf I-\overline{\mathbf{u}}_{1}\overline{\mathbf{u}}_{1}^\top)\widetilde{\mathbf{C}}\\
        &=\frac{1}{2d\gamma n}\widetilde{\mathbf{C}}^{\top}\widetilde{\mathbf{B}}^2\widetilde{\mathbf{C}}.
    \end{align*}
 which completes the proof. \hfill $\square$

\emph{Proof of Proposition \ref{Corollary-star}. }
\par 

From Lemma \ref{lemma_eigenvalues} and the assumption of $m_i=\eta$ for all the nodes and that of $K_{i,j}\cos\delta_{ij}^*=\gamma$ for all the lines, we obtain the eigenvalues of 
the matrix $\mathbf{M}^{-1/2}\mathbf{L}\mathbf{M}^{-1/2}$ as defined in (\ref{decomposition}),
\begin{align*}
\lambda_1=0,\lambda_i=\gamma\eta^{-1}~\text{for}~i=2,\cdots, n~\text{and}~\lambda_n=\gamma\eta^{-1} n. 
\end{align*}
Because the vector $\begin{bmatrix}
n-1&-1&-1&\cdots&-1
\end{bmatrix}^\top$ is the eigenvector corresponding to the eigenvalue $\lambda_n$,   we obtain $\bm u_n=1/\sqrt{n(n-1)}\begin{bmatrix}
n-1&-1&-1&\cdots&-1
\end{bmatrix}^\top$. 
Denote $\widehat{\mathbf{U}}=[\widehat{\mathbf{U}}_2 ~~\mathbf{u}_n]$, where $\widehat{\mathbf{U}}_2\in\mathbb{R}^{n\times (n-2)}$. Let $\rho=\frac{\alpha(1+n)}{\eta^{-1}\gamma(n-1)^2+2\alpha^2(1+n)}$, we obtain 
the formula of the matrix $\mathbf{R}$ from Theorem \ref{maintheorem_kuramot}, 
\begin{align*}
    \mathbf{R}=\eta^{-1}
    \begin{bmatrix}
        \frac{1}{2\alpha}\mathbf{u}_1^\top \widetilde{\mathbf{B}}^2\mathbf{u}_1 & \frac{\alpha}{2\alpha^2+\eta^{-1}\gamma}\mathbf{u}_1^\top\widetilde{\mathbf{B}}^2\widehat{\mathbf{U}}_2 & \frac{\alpha}{2\alpha^2+\eta^{-1}\gamma n}\mathbf{ u}_1^\top\widetilde{\mathbf{B}}^2\widehat{\mathbf{u}}_n\\
        \frac{\alpha}{2\alpha^2+\eta^{-1}\gamma}\widehat{\mathbf{U}}_2^\top\widetilde{\mathbf{B}}^2\mathbf{u}_1 & \frac{1}{2\alpha}\widehat{\mathbf{U}}_2^\top \widetilde{\mathbf{B}}^2\widehat{\mathbf{U}}_2 & \rho\widehat{\mathbf{U}}_2^\top\widetilde{\mathbf{B}}^2\mathbf{u}_n\\
        \frac{\alpha}{2\alpha^2+\eta^{-1}\gamma n}\mathbf{u}_n^\top\widetilde{\mathbf{B}}^2\mathbf{u}_1 &
        \rho\mathbf{u}_n^\top\widetilde{\mathbf{B}}^2\widehat{\mathbf{U}}_2 & \frac{1}{2\alpha}\mathbf{u}_n^\top\widetilde{\mathbf{B}}^2\mathbf{u}_n
    \end{bmatrix}
\end{align*}
Thus, the variance matrix $\mathbf{Q}_\omega$ becomes
\begin{align*}
    \mathbf{Q}_{\bm\omega}&=\eta^{-2}
    \begin{bmatrix}
        \mathbf{u}_1 & \widehat{\mathbf{U}}_2 &\mathbf{u}_n
    \end{bmatrix}
    \mathbf{R}
    \begin{bmatrix}
        \mathbf{u}_1 & \widehat{\mathbf{U}}_2 &\mathbf{u}_n
    \end{bmatrix}^\top\\
    &\text{by} ~\mathbf{u}_1\mathbf{u}_1^\top+\widehat{\mathbf{U}}_2\widehat{\mathbf{U}}_2^\top+\mathbf{u}_n\mathbf{u}_n^\top=\mathbf{I}\\
     &=\eta^{-2}
    \left(
        \frac{1}{2\alpha}\widetilde{\mathbf{B}}^2+
        \frac{\gamma\left(
            2\mathbf{u}_1\mathbf{u}_1^\top\widetilde{\mathbf{B}}^2\mathbf{u}_1\mathbf{u}_1^\top
        -\mathbf{u}_1\mathbf{u}_1^\top\widetilde{\mathbf{B}}^2
        -\widetilde{\mathbf{B}}^2\mathbf{u}_1\mathbf{u}_1^\top
        \right)}{2\alpha\eta(\eta^{-1}\gamma+2\alpha^2)}\right.\\
        &~~+
        \frac{\gamma(n-1)^2\left(
            2\mathbf{u}_n\mathbf{u}_n^\top\widetilde{\mathbf{B}}^2\mathbf{u}_n\mathbf{u}_n^\top
            -\mathbf{u}_n\mathbf{u}_n^\top\widetilde{\mathbf{B}}^2
            -\widetilde{\mathbf{B}}^2\mathbf{u}_n\mathbf{u}_n^\top
        \right)}{2\alpha\eta\left(\eta^{-1}\gamma(n-1)^2+2\alpha^2(1+n)\right)}\\
        &~~+
        \frac{\gamma(n-1)\left(\eta^{-2}\gamma^2 n(n-1)+4\alpha^2\eta^{-1}\gamma(n-1)-8\alpha^4\right)}{2\alpha\eta(\eta^{-1}\gamma+2\alpha^2)(\eta^{-1}\gamma n+2\alpha^2)\left(\eta^{-1}\gamma(n-1)^2+2\alpha^2(1+n)\right)}\\
        &~~
        \times\left(
            \mathbf{u}_1\mathbf{u}_1^\top\widetilde{\mathbf{B}}^2\mathbf{u}_n\mathbf{u}_n^\top+
            \mathbf{u}_n\mathbf{u}_n^\top\widetilde{\mathbf{B}}^2\mathbf{u}_1\mathbf{u}_1^\top
        \right)
    \Bigg)
\end{align*}
With the explicit formulas of $\mathbf{u}_1$ and $\mathbf{u}_n$,  we obtain the entries of the matrices, 
\begin{align*}
    &\mathbf{u}_1\mathbf{u}_1^\top\widetilde{\mathbf{B}}^2:\frac{1}{n}b_j^2,\\
    &\widetilde{\mathbf{B}}^2\mathbf{u}_1\mathbf{u}_1^\top:\frac{1}{n}b_i^2,\\
    &\mathbf{u}_1\mathbf{u}_1^\top\widetilde{\mathbf{B}}^2\mathbf{u}_1\mathbf{u}_1^\top:\frac{1}{n^2}\text{tr}\left(\widetilde{\mathbf{B}}^2\right),\\
    &\mathbf{u}_n\mathbf{u}_n^\top\widetilde{\mathbf{B}}^2:\frac{1}{n(n-1)}
    \begin{cases}
        ~~(1-n)^2 b_1^2,~~ i\!=\!j\!=\!1,\\
        ~~(1-n) b_1^2,~~i\!=\!2,\cdots,n,j\!=\!1,\\
        ~~(1-n) b_j^2,~~j=2,\cdots,n,i\!=\!1,\\
        ~~ b_j^2,~~i,j\!=\!2,\cdots,n,
    \end{cases}\\
    &\widetilde{\mathbf{B}}^2\mathbf{u}_n\mathbf{u}_n^\top:\frac{1}{n(n-1)}
    \begin{cases}
        ~~(1-n)^2 b_1^2,~~ i\!=\!j\!=\!1,\\
        ~~(1-n) b_1^2,~~j\!=\!2,\cdots,n,i\!=\!1,\\
        ~~(1-n) b_i^2,~~i\!=\!2,\cdots,n,j=1,\\
        ~~ b_i^2,~~i,j\!=\!2,\cdots,n,
    \end{cases}\\
    &\mathbf{u}_1\mathbf{u}_1^\top\widetilde{\mathbf{B}}^2\mathbf{u}_n\mathbf{u}_n^\top:\frac{1}{n^2(n-1)}
    \begin{cases}
        ~~(1-n)^2 b_1^2+(1-n)\sum\limits_{t=2}^n b_t^2,~~j\!=\!1,\\
        ~~ (1-n) b_1^2+\sum\limits_{t=2}^n b_t^2,~~\text{otherwise},
    \end{cases}\\
    &\mathbf{u}_n\mathbf{u}_n^\top\widetilde{\mathbf{B}}^2\mathbf{u}_1\mathbf{u}_1^\top:\frac{1}{n^2(n-1)}
    \begin{cases}
        ~~(1-n)^2 b_1^2+(1-n)\sum\limits_{t=2}^n b_t^2,~~i\!=\!1,\\
        ~~ (1-n) b_1^2+\sum\limits_{t=2}^n b_t^2,~~\text{otherwise},
    \end{cases}\\
    &\mathbf{u}_n\mathbf{u}_n^\top\widetilde{\mathbf{B}}^2\mathbf{u}_n\mathbf{u}_n^\top:\frac{1}{n^2(n-1)^2}
    \begin{cases}
        ~~(1-n)^4 b_1^2+(1-n)^2\sum\limits_{t=2}^n b_t^2,~~i\!=\!j\!=\!1,\\
        ~~ (1-n)^2 b_1^2+\sum\limits_{t=2}^n b_t^2,~~i,j\!=\!2,\!\cdots\!,n,\\
        ~~(1-n)^3 b_1^2+(1-n)\sum\limits_{t=2}^n b_t^2,~\text{otherwise}.
    \end{cases}         
\end{align*}
where $i,j$ represent $i$th row and $j$th column in left matrices respectively. With these equations,
we get
\begin{align*}
    q_{\omega_{1,1}}=&\eta^{-2}
    \left(
        \frac{1}{2\alpha}b_1^2+
        \frac{\gamma\left(
            \text{tr}\left(\widetilde{\mathbf{B}}^2\right)
        -n b_1^2
        \right)}{\alpha\eta n^2(\eta^{-1}\gamma+2\alpha^2)}\right.\\
        &+
        \frac{\gamma(n-1)^2\left(
            (1-n)^2 b_1^2+\sum_{t=2}^n b_t^2
            -(n-1)n b_1^2
        \right)}{\alpha\eta n^2\left(\eta^{-1}\gamma(n-1)^2+2\alpha^2(1+n)\right)}\\
        &+
        \frac{\gamma(n-1)\left(\eta^{-2}\gamma^2 n(n-1)+4\alpha^2\eta^{-1}\gamma(n-1)-8\alpha^4\right)}{\alpha\eta n^2(\eta^{-1}\gamma+2\alpha^2)(\eta^{-1}\gamma n+2\alpha^2)\left(\eta^{-1}\gamma(n-1)^2+2\alpha^2(1+n)\right)}\\
        &
        \times\Big(
            (n-1) b_1^2-\sum_{t=2}^n b_t^2
        \Big)
    \Bigg)\\
    &=\big[\frac{1}{2d\eta}-\frac{\gamma(n-1)}{dn(2d^2+\gamma\eta n)}\big]b_{1}^2
    +\frac{\gamma}{d n\left(2d^2+\gamma \eta n\right)}(\text{tr}(\widetilde{\mathbf{B}}^2)-b_{1}^2)
\end{align*}
and for $i=2,\cdots,n$,
\begin{align*}
    q_{\omega_{i,i}}=&\eta^{-2}
    \left(
        \frac{1}{2\alpha}b_i^2+
        \frac{\gamma\left(
            \text{tr}\left(\widetilde{\mathbf{B}}^2\right)-n b_i^2
        \right)}{\alpha\eta n^2(\eta^{-1}\gamma+2\alpha^2)}\right.
        +
        \frac{\gamma\left(
            (1-n)^2 b_1^2+\sum_{t=2}^n b_t^2
            -n(n-1)b_i^2
        \right)}{\alpha\eta n^2\left(\eta^{-1}\gamma(n-1)^2+2\alpha^2(1+n)\right)}\\
        &+
        \frac{\gamma\left(\eta^{-2}\gamma^2 n(n-1)+4\alpha^2\eta^{-1}\gamma(n-1)-8\alpha^4\right)}{\alpha\eta n^2(\eta^{-1}\gamma+2\alpha^2)(\eta^{-1}\gamma n+2\alpha^2)\left(\eta^{-1}\gamma(n-1)^2+2\alpha^2(1+n)\right)}\\
        &
        \times\Big(
            (1-n)b_1^2+\sum_{t=2}^n b_t^2
        \Big)
    \Bigg)\\
    =&
       \frac{\gamma}{d n\left(2d^2+\gamma \eta n\right)}b_{1}^2+\frac{1}{2d\eta}b_i^2-\frac{\gamma}{dn(2d^2+\gamma\eta n)}b_i^2\\
        &-\frac{\gamma(n-2)b_i^2}{dn(2d^2(n+1)+\gamma\eta(n-1)^2)}-\frac{\gamma^2\eta(n-2)b_i^2}{dn(2d^2+\gamma\eta)(2d^2+\gamma\eta n)}\\       
        &+\frac{\gamma}{dn(2d^2(1\!+\!n)\!+\!\gamma\eta (n-1)^2)}(\text{tr}(\widetilde{\mathbf{B}}^2)\!-\!b_i^2\!-\!b_1^2)\\
        &+\frac{\gamma^2\eta}{dn(2d^2\!+\!\gamma\eta)(2d^2\!+\!\gamma\eta n)}(\text{tr}(\widetilde{\mathbf{B}}^2)\!-\!b_i^2\!-\!b_1^2).
\end{align*}
Now, we calculate the variance of the phase difference. With the explicit 
formulas of the eigenvalues $\lambda_i$,  let $\epsilon =\frac{2\alpha}{2\alpha^2\eta^{-1}\gamma(1+n)+\eta^{-2}\gamma^2(n-1)^2} $,  we obtain
\begin{align*}
    \mathbf{G}=\eta^{-1}
    \begin{bmatrix}
        \frac{1}{2\alpha\eta^{-1}\gamma}\widehat{\mathbf{U}}_2^\top\widetilde{\mathbf{B}}^2\widehat{\mathbf{U}}_2& \epsilon  \widehat{\mathbf{U}}_2^\top \widetilde{\mathbf{B}}^2\mathbf{u}_n\\
        \epsilon  \bm u_n^\top \widetilde{\mathbf{B}}^2\widehat{\mathbf{U}}_2 & \frac{1}{2\alpha\eta^{-1}\gamma n}\mathbf{u}_n^\top\widetilde{\mathbf{B}}^2\mathbf{u}_n
    \end{bmatrix}
\end{align*}
Let $\mathbf{T}=\mathbf{M}^{-\frac{1}{2}}\widehat{\mathbf{U}}\mathbf{G}\widehat{\mathbf{U}}^\top\mathbf{M}^{-\frac{1}{2}}$ , we get
\begin{align*}
    &\mathbf{T}=\eta^{-2}
    \begin{bmatrix}
        \widehat{\mathbf{U}}_2 &\mathbf{u}_n
    \end{bmatrix}
    \mathbf{G}
    \begin{bmatrix}
        \widehat{\mathbf{U}}_2 &\mathbf{u}_n
    \end{bmatrix}^\top\\
  &\text{by} ~\mathbf{u}_1\mathbf{u}_1^\top+\widehat{\mathbf{U}}_2\widehat{\mathbf{U}_2}^\top+\mathbf{u}_n\mathbf{u}_n^\top=\mathbf{I}\\
    &\!=\eta^{-2}
    \left(
        \frac{1}{2\alpha\eta^{-1}\gamma}\widetilde{\mathbf{B}}^2+\frac{\mathbf{u}_1\mathbf{u}_1^\top\widetilde{\mathbf{B}}^2\mathbf{u}_1\mathbf{u}_1^\top
        -\mathbf{u}_1\mathbf{u}_1^\top\widetilde{\mathbf{B}}^2
        -\widetilde{\mathbf{B}}^2\mathbf{u}_1\mathbf{u}_1^\top}{2\alpha\eta^{-1}\gamma}\!\right.\\
        &+
        \!\frac{(\eta^{-1}\gamma(n-1)^2+2\alpha^2(n-1))\big(\mathbf{u}_1\mathbf{u}_1^\top\widetilde{\mathbf{B}}^2\mathbf{u}_n\mathbf{u}_n^\top\!+\!\mathbf{u}_n\mathbf{ u}_n^\top\widetilde{\mathbf{B}}^2\mathbf{u}_1\mathbf{u}_1^\top
        \!-\!\mathbf{u}_n\mathbf{u}_n^\top\widetilde{\mathbf{B}}^2
        \!-\!\widetilde{\mathbf{B}}^2\mathbf{u}_n\mathbf{u}_n^\top\big)}{2\alpha\eta^{-1}\gamma (2\alpha^2(1+n)+\eta^{-1}\gamma(n-1)^2)}\!\\
        &
    \\
    &+\left.
        \frac{\eta^{-1}\gamma(n+1)(n-1)^2+2\alpha^2(n-1)^2}{2\alpha\eta^{-1}\gamma n(2\alpha^2(1+n)+\eta^{-1}\gamma(n-1)^2)}\mathbf{u}_n\mathbf{u}_n^\top\widetilde{\mathbf{B}}^2\mathbf{u}_n\mathbf{u}_n^\top
    \right)
\end{align*}
With the form of the incidence matrix of the star graph in Lemma \ref{lemma_eigenvalues},  it yields
 from (\ref{eq:variance0}) that 
        \begin{equation*}
            \begin{aligned}
            q_{\delta_{k,q}}
            =
            T_{11}-T_{k+1,1}-T_{1,q+1}+T_{k+1,q+1}
            \end{aligned}
        \end{equation*}
  the formulas of $T_{11},T_{k+1,1},T_{1,q+1},T_{k+1,q+1}$ as follows,
    \begin{align*}
        T_{11}=&\eta^{-2}
        \Bigg(
        \frac{1}{2\alpha\eta^{-1}\gamma} b_1^2-\frac{2b_1^2}{2\alpha\eta^{-1}\gamma n}+\frac{1}{2\alpha\eta^{-1}\gamma n^2}\text{tr}(\widetilde{\mathbf{B}}^2)\\
        &+ \frac{2(\eta^{-1}\gamma(n-1)+2\alpha^2)\left(
            (1-n)^2 b_1^2+(1-n)\sum_{t=2}^n b_t^2
        \right)}{2\alpha\eta^{-1}\gamma n^2 (2\alpha^2(1+n)+\eta^{-1}\gamma(n-1)^2)}\\
        &-
        \frac{2(1-n)^2(\eta^{-1}\gamma(n-1)+2\alpha^2) b_1^2}{2\alpha\eta^{-1}\gamma n (2\alpha^2(1+n)+\eta^{-1}\gamma(n-1)^2)}\\
        &+
        \frac{(\eta^{-1}\gamma(n+1)+2\alpha^2)\left(
            (1-n)^4 b_1^2+(1-n)^2\sum_{t=2}^n b_t^2
        \right)}{2\alpha\eta^{-1}\gamma n^3(2\alpha^2(1+n)+\eta^{-1}\gamma(n-1)^2)}
        \Bigg),
    \end{align*}
    \begin{align*}
        T_{k+1,1}=&\eta^{-2}
        \Bigg(
        -\frac{b_1^2+b_{k+1}^2}{2\alpha\eta^{-1}\gamma n}+\frac{1}{2\alpha\eta^{-1}\gamma n^2}\text{tr}(\widetilde{\mathbf{B}}^2)\\
        &+ \frac{\eta^{-1}\gamma(n-1)+2\alpha^2}{2\alpha\eta^{-1}\gamma n^2 (2\alpha^2(1+n)+\eta^{-1}\gamma(n-1)^2)}\\
        &\times\Big(
            (1-n)^2 b_1^2+(1-n)\sum_{t=2}^n b_t^2
            +
            (1-n) b_1^2+\sum_{t=2}^n b_t^2
        \Big)\\
        &-
        \frac{(1-n)(\eta^{-1}\gamma(n-1)+2\alpha^2)(b_1^2+b_{k+1}^2)}{2\alpha\eta^{-1}\gamma n (2\alpha^2(1+n)+\eta^{-1}\gamma(n-1)^2)}\\
        &+
        \frac{(\eta^{-1}\gamma(n+1)+2\alpha^2)\left(
            (1-n)^3 b_1^2+(1-n)\sum_{t=2}^n b_t^2
        \right)}{2\alpha\eta^{-1}\gamma n^3(2\alpha^2(1+n)+\eta^{-1}\gamma(n-1)^2)}
        \Bigg),
    \end{align*}
    \begin{align*}
        T_{1,q+1}=&\eta^{-2}
        \Bigg(
        -\frac{b_1^2+b_{q+1}^2}{2\alpha\eta^{-1}\gamma n}+\frac{1}{2\alpha\eta^{-1}\gamma n^2}\text{tr}(\widetilde{\mathbf{B}}^2)\\
        &+ \frac{\eta^{-1}\gamma(n-1)+2\alpha^2}{2\alpha\eta^{-1}\gamma n^2 (2\alpha^2(1+n)+\eta^{-1}\gamma(n-1)^2)}\\
        &\times\Big(
            (1-n)^2 b_1^2+(1-n)\sum_{t=2}^n b_t^2
            +
            (1-n) b_1^2+\sum_{t=2}^n b_t^2
        \Big)\\
        &-
        \frac{(1-n)(\eta^{-1}\gamma(n-1)+2\alpha^2)(b_1^2+b_{q+1}^2)}{2\alpha\eta^{-1}\gamma n (2\alpha^2(1+n)+\eta^{-1}\gamma(n-1)^2)}\\
        &+
        \frac{(\eta^{-1}\gamma(n+1)+2\alpha^2)\left(
            (1-n)^3 b_1^2+(1-n)\sum_{t=2}^n b_t^2
        \right)}{2\alpha\eta^{-1}\gamma n^3(2\alpha^2(1+n)+\eta^{-1}\gamma(n-1)^2)}
        \Bigg),
    \end{align*}
    If $k\neq q$, we have
    \begin{align*}
        T_{k+1,q+1}=&\eta^{-2}
        \Bigg(
        -\frac{b_{k+1}^2+b_{q+1}^2}{2\alpha\eta^{-1}\gamma n}+\frac{1}{2\alpha\eta^{-1}\gamma n^2}\text{tr}
       ( \widetilde{\mathbf{B}}^2)\\
        &+ \frac{2(\eta^{-1}\gamma(n-1)+2\alpha^2)\left(
            (1-n) b_1^2+\sum_{t=2}^n b_t^2
        \right)}{2\alpha\eta^{-1}\gamma n^2 (2\alpha^2(1+n)+\eta^{-1}\gamma(n-1)^2)}\\
        &-
        \frac{(\eta^{-1}\gamma(n-1)+2\alpha^2)(b_{k+1}^2+b_{q+1}^2)}{2\alpha\eta^{-1}\gamma n (2\alpha^2(1+n)+\eta^{-1}\gamma(n-1)^2)}\\
        &+
        \frac{(\eta^{-1}\gamma(n+1)+2\alpha^2)\left(
            (1-n)^2 b_1^2+\sum_{t=2}^n b_t^2
        \right)}{2\alpha\eta^{-1}\gamma n^3(2\alpha^2(1+n)+\eta^{-1}\gamma(n-1)^2)}
        \Bigg),
    \end{align*}
    Then
    \begin{align*}
        q_{\delta_{k,q}}
        =&\frac{1}{2d\gamma n} b_{1}^2+
        \frac{-2d^2(n-1)+\gamma\eta(2n-n^2+1)}{2d\gamma n(2d^2(1+n)+\gamma\eta(n-1)^2)} b_{k+1}^2\\
        &~+
        \frac{-2d^2(n-1)+\gamma\eta(2n-n^2+1)}{2d\gamma n(2d^2(1+n)+\gamma\eta(n-1)^2)} b_{q+1}^2\\
        &
        ~+\frac{2d^2+\gamma\eta(n+1)}{2d\gamma n(2d^2(1+n)+\gamma\eta(n-1)^2)} \left(\text{tr}(\widetilde{\mathbf{B}}^2)-b_{k+1}^2-b_{q+1}^2-b_1^2\right)    
    \end{align*}
    If $k=q$, we have
    \begin{align*}
        T_{k+1,k+1}=&\eta^{-2}
        \Bigg(
        \frac{1}{2\alpha\eta^{-1}\gamma} b_{k+1}^2-\frac{2b_{k+1}^2}{2\alpha\eta^{-1}\gamma n}+\frac{1}{2\alpha\eta^{-1}\gamma n^2}\text{tr}(\widetilde{\mathbf{B}}^2)\\
        &+ \frac{2(\eta^{-1}\gamma(n-1)+2\alpha^2)\left(
            (1-n) b_1^2+\sum_{t=2}^n b_t^2
        \right)}{2\alpha\eta^{-1}\gamma n^2 (2\alpha^2(1+n)+\eta^{-1}\gamma(n-1)^2)}\\
        &-
        \frac{2(\eta^{-1}\gamma(n-1)+2\alpha^2)b_{k+1}^2}{2\alpha\eta^{-1}\gamma n (2\alpha^2(1+n)+\eta^{-1}\gamma(n-1)^2)}\\
        &+
        \frac{(\eta^{-1}\gamma(n+1)+2\alpha^2)\left(
            (1-n)^2 b_1^2+\sum_{t=2}^n b_t^2
        \right)}{2\alpha\eta^{-1}\gamma n^3(2\alpha^2(1+n)+\eta^{-1}\gamma(n-1)^2)}
        \Bigg),
    \end{align*}
    Then we substitute it into $q_{\delta_{k,k}}$ to get
    \begin{align*}
    q_{\delta_{k,k}}
    =&\frac{1}{2d\gamma n} b_{1}^2+\big(\frac{n-1}{2d\gamma n}-
    \frac{(n-2)(2d^2+\gamma\eta(n+1))}{2d\gamma n(2d^2(1+n)+\gamma\eta(n-1)^2)}\big) b_{k+1}^2\\\
    &+\frac{2d^2+\gamma\eta(n+1)}{2d\gamma n(2d^2(1+n)+\gamma\eta(n-1)^2)} \left(\text{tr}(\widetilde{\mathbf{B}}^2)-b_{k+1}^2-b_1^2\right)
    \end{align*}
    Then we complete the proof. \hfill $\square$
    \par 
 \emph{Proof of Proposition \ref{proposition_kuramoto}}.  
 \par 
Following Lemma \ref{lemma_eigenvalues}(ii) and the assumption of $\mathbf{D}=d\mathbf{I}$
and $l_{c_{i,j}}=\gamma$ for all the lines, we obtain the eigenvalues of 
the matrix $\mathbf{D}^{-1/2}\mathbf{L}\mathbf{D}^{-1/2}$, 
\begin{align*}
\overline{\lambda}_1=0,~~\overline{\lambda}_2=\cdots=\overline{\lambda}_{n-1}=\gamma/d, ~~\overline{\lambda}_n=n\gamma/d~\text{for}~i=2,\cdots,n. 
\end{align*} 
 Following 
the formula of the matrix $\overline{\mathbf{Q}}_x$ in (\ref{Q1}), we obtain, 
    \begin{equation}
        \overline{q}_{x_{i,j}}=
        \begin{cases}
            \frac{1}{2\gamma}\overline{\mathbf{u}}_i^\top \widetilde{\mathbf{B}}^2\overline{\mathbf{u}}_j, & i,j=2,\cdots,n-1,\\
            \frac{1}{\gamma(1+n)}\overline{\mathbf{u}}_n^\top \widetilde{\mathbf{B}}^2\overline{\mathbf{u}}_j, & i=n, j=2,\cdots,n-1,\\
            \frac{1}{\gamma(1+n)}\overline{\mathbf{u}}_i^\top \widetilde{\mathbf{B}}^2\overline{\mathbf{u}}_n, & j=n, i=2,\cdots,n-1,\\
            \frac{1}{2\gamma n} \overline{\mathbf{u}}_n^\top \widetilde{\mathbf{B}}^2\overline{\mathbf{u}}_n, & i=j=n.
        \end{cases}
    \end{equation}
Denote $\overline{\mathbf{U}}_2=[\widehat{\overline{\mathbf{U}}}_2~~ \overline{\mathbf{u}}_n]$, where $\widehat{\overline{\mathbf{U}}}_2\in\mathbb{R}^{n\times (n-2)}$. Then we convert the matrix $\overline{\mathbf{Q}}_{x}$ into four blocks,
    \begin{equation*}
        \overline{\mathbf{Q}}_{x}=
        \begin{bmatrix}
            \frac{1}{2\gamma}\widehat{\overline{\mathbf{U}}}_2^\top\widetilde{\mathbf{B}}^2\widehat{\overline{\mathbf{U}}}_2 & \frac{1}{\gamma (1+n)}\widehat{\overline{\mathbf{U}}}_2^\top\widetilde{\mathbf{B}}^2\overline{\mathbf{u}}_n\\
            \frac{1}{\gamma (1+n)}\overline{\mathbf{u}}_n^\top\widetilde{\mathbf{B}}^2\widehat{\overline{\mathbf{U}}}_2 & \frac{1}{2\gamma n}\overline{\mathbf{u}}_n^\top\widetilde{\mathbf{B}}^2\overline{\mathbf{u}}_n
        \end{bmatrix}
    \end{equation*}
    Let $\widetilde{\mathbf{T}}=\overline{\mathbf{U}}_2
    \overline{\mathbf{Q}}_{x}
 \overline{\mathbf{U}}_2^{\top}$, then we have
 \begin{align*}
    &\widetilde{\mathbf{T}}=[\widehat{\overline{\mathbf{U}}}_2~~\overline{\mathbf{u}}_n]\begin{bmatrix}
        \frac{1}{2\gamma}\widehat{\overline{\mathbf{U}}}_2^\top\widetilde{\mathbf{B}}^2\widehat{\overline{\mathbf{U}}}_2 & \frac{1}{\gamma (1+n)}\widehat{\overline{\mathbf{U}}}_2^\top\widetilde{\mathbf{B}}^2\overline{\mathbf{u}}_n\widetilde{\mathbf{C}}\\
        \frac{1}{\gamma (1+n)}\overline{\mathbf{u}}_n^\top\widetilde{\mathbf{B}}^2\widehat{\overline{\mathbf{U}}}_2 & \frac{1}{2\gamma n}\overline{\mathbf{u}}_n^\top\widetilde{\mathbf{B}}^2\overline{\mathbf{u}}_n
    \end{bmatrix}[\widehat{\overline{\mathbf{U}}}_2~~\overline{\mathbf{u}}_n]^\top\\
    &=\frac{1}{2\gamma}\widetilde{\mathbf{B}}^2+\frac{1}{2\gamma}\left(\overline{\mathbf{u}}_1\overline{\mathbf{u}}_1^\top\widetilde{\mathbf{B}}^2\overline{\mathbf{u}}_1\overline{\mathbf{u}}_1^\top-\overline{\mathbf{u}}_1\overline{\mathbf{u}}_1^\top\widetilde{\mathbf{B}}^2-\widetilde{\mathbf{B}}^2\overline{\mathbf{u}}_1\overline{\mathbf{u}}_1^\top\right)+\frac{(n-1)^2}{2\gamma n(1+n)}\overline{\mathbf{u}}_n\overline{\mathbf{u}}_n^\top\widetilde{\mathbf{B}}^2\overline{\mathbf{u}}_n\overline{\mathbf{u}}_n^\top\\
    &+\frac{n-1}{2\gamma(1+n)}\left(\overline{\mathbf{u}}_1\overline{\mathbf{u}}_1^\top\widetilde{\mathbf{B}}^2\overline{\mathbf{u}}_n\overline{\mathbf{u}}_n^\top+\overline{\mathbf{u}}_n\overline{\mathbf{u}}_n^\top\widetilde{\mathbf{B}}^2\overline{\mathbf{u}}_1\overline{\mathbf{u}}_1^\top-\widetilde{\mathbf{B}}^2\overline{\mathbf{u}}_n\overline{\mathbf{u}}_n^\top-\overline{\mathbf{u}}_n\overline{\mathbf{u}}_n^\top\widetilde{\mathbf{B}}^2\right)
 \end{align*}
So we get the formula of the variance matrix of the phase angle differences,
    \begin{align*}
        \overline{\mathbf{Q}}_\delta=
        \frac{1}{d}\widetilde{\mathbf{C}}^{\top}\widetilde{\mathbf{T}}\widetilde{\mathbf{C}}
    \end{align*}
    and substitute the incidence matrix
    $\widetilde{\mathbf{C}}$ into the above equation, we have
    \begin{equation*}
        \begin{aligned}
        \overline{q}_{\delta_{k,q}}
        =\frac{1}{d}\left(\widetilde{T}_{11}-\widetilde{T}_{k+1,1}-\widetilde{T}_{1,q+1}+\widetilde{T}_{k+1,q+1}\right).
        \end{aligned}
    \end{equation*}
   Following Lemma \ref{lemma_eigenvalues}(ii), $\overline{\mathbf{u}}_n=1/\sqrt{n(n-1)}[1-n~~1~~\cdots~~1]^\top$ is the eigenvector corresponding to the eigenvalue $n$, then
    \begin{align*}
        \widetilde{T}_{11}=&\frac{1}{2\gamma}b_1^2+\frac{1}{2\gamma n^3(1+n)}\left[(1-n)^4 b_1^2+(1-n)^2\sum\limits_{t=2}^n b_t^2\right]+\frac{1}{2\gamma}\left(\frac{1}{n^2}\text{tr}\left(\widetilde{\mathbf{B}}^2\right)-\frac{2}{n}b_1^2\right)\\
        &+\frac{1}{2\gamma n^2(1+n)}\left(2(1-n)^2b_1^2+2(1-n)\sum_{t=2}^n b_t^2-2n(1-n)^2b_1^2\right),\\
        \widetilde{T}_{1,q+1}=&\frac{1}{2\gamma n^3(1+n)}\left[(1-n)^3 b_1^2+(1-n)\sum\limits_{t=2}^n b_t^2\right]+\frac{1}{2\gamma}\left(\frac{1}{n^2}\text{tr}\left(\widetilde{\mathbf{B}}^2\right)-\frac{b_1^2+b_{q+1}^2}{n}\right)\\
        &\!+\frac{1}{2\gamma n^2(1+n)}\left((2-n)(1-n)b_1^2+(2-n)\sum_{t=2}^n b_t^2-n(1-n)b_1^2-n(1-n)b_{q+1}^2\right),\!\\
        \widetilde{T}_{k+1,1}=&\frac{1}{2\gamma n^3(1+n)}\left[(1-n)^3 b_1^2+(1-n)\sum\limits_{t=2}^n b_t^2\right]+\frac{1}{2\gamma}\left(\frac{1}{n^2}\text{tr}\left(\widetilde{\mathbf{B}}^2\right)-\frac{b_1^2+b_{k+1}^2}{n}\right)\\
        &\!+\frac{1}{2\gamma n^2(1+n)}\left((2-n)(1-n)b_1^2+(1-n)\sum_{t=2}^n b_t^2+(2-n) b_1^2-n(1-n)b_{k+1}^2\right)\!,\\
        \widetilde{T}_{k+1,q+1}=&\frac{1}{2\gamma n^3(1+n)}\left[(1-n)^2 b_1^2+\sum\limits_{t=2}^n b_t^2\right]+\frac{1}{2\gamma}\left(\frac{1}{n^2}\text{tr}\left(\widetilde{\mathbf{B}}^2\right)-\frac{b_{q+1}^2+b_{k+1}^2}{n}\right)\\
        &\!+\frac{1}{2\gamma n^2(1+n)}\left(2(1-n) b_1^2+2\sum_{t=2}^n b_t^2-n(b_{k+1}^2+b_{q+1}^2)\right),\, (k\neq q),\\
        \widetilde{T}_{k+1,k+1}=&\frac{1}{2\gamma}b_{k+1}^2+\frac{1}{2\gamma n^3(1+n)}\left[(1-n)^2 b_1^2+\sum\limits_{t=2}^n b_t^2\right]\\
        &\!+\frac{1}{2\gamma n^2(1+n)}\left(2(1-n) b_1^2+2\sum_{t=2}^n b_t^2-2nb_{k+1}^2\right)+\frac{1}{2\gamma}\left(\frac{1}{n^2}\text{tr}\left(\widetilde{\mathbf{B}}^2\right)-\frac{2b_{k+1}^2}{n}\right)\!.
    \end{align*}
    We further obtain
    \begin{align*}
        \overline{q}_{\delta_{k,q}}
        =\frac{1}{d}\left(\widetilde{T}_{11}-\widetilde{T}_{k+1,1}-\widetilde{T}_{1,q+1}+\widetilde{T}_{k+1,q+1}\right)
    \end{align*}
   and
    \begin{align*}
        \overline{q}_{\delta_{k,k}}
        =\frac{1}{d}\left(\widetilde{T}_{11}-\widetilde{T}_{k+1,1}-\widetilde{T}_{1,k+1}+\widetilde{T}_{k+1,k+1}\right)
    \end{align*}
    which leads to (\ref{overlineQ0}) and (\ref{overlineQ1}) respectively. 
\hfill $\square$
\section{Conclusions}
\label{sec:conclusions}
The analytic formula of the variance matrix of a stochastic system linearized from a power system has been deduced at the invariant probability distribution based on the assumption of uniform damping-inertia ratio at all nodes. With this analytic formula and assumption of 
identical weights of the lines, the impact of the system parameters on the propagation of the fluctuations in the system with complete graphs and 
star graphs is analyzed. 
\par
Research interest remains for the analytic formula of the variance matrix without any assumptions
on the system parameters. 
%

\bibliography{sample,base,basetmp}
\bibliographystyle{siamplain}
\end{document}

%% file: syskuramotovar-shared.tex
\usepackage{amsfonts}
\usepackage{bm}
\usepackage{enumerate}
\usepackage{multirow}
\usepackage{graphicx}
\usepackage{epstopdf}
\usepackage{algorithmic}
\ifpdf
  \DeclareGraphicsExtensions{.eps,.pdf,.png,.jpg}
\else
  \DeclareGraphicsExtensions{.eps}
\fi

\input{macros2e}

\newsiamthm{problem}{Problem}
{\theorembodyfont{\rmfamily} }

\newsiamthm{claim}{Claim}
\newsiamremark{hypothesis}{Hypothesis}
\crefname{hypothesis}{Hypothesis}{Hypotheses}
\newsiamremark{remark}{Remark}

\headers{The variance of a stochastic process for power systems}{Xian Wu and Kaihua Xi {et al}}

\title{
Explicit formulas for the Variance of the State \\of a Linearized Power System\\
driven by Gaussian stochastic disturbances
   \thanks{The authors thank Mr. Zhen Wang of Shandong University for 
the useful discussions and comments on this research topic.
    }
}

\author{
~~~~~~Xian Wu
\thanks{School of Mathematics, Shandong University, Jinan, 250100, 
    Shandong Province, China%
    (\email{kxi@sdu.edu.cn}%
    )%
  }
\and
Kaihua Xi
  \footnotemark[2]
\and
Aijie Cheng
  \footnotemark[2]
\and
Hai Xiang Lin
\thanks{Delft Institute of Applied Mathematics, Delft University of Technology,
    Delft, 2628CD, The Netherlands%
  }
\and 
\newline
Jan H. van Schuppen
\footnotemark[3]
\and
Chenghui Zhang
\thanks{School of Control Science and Engineering, Shandong University, Jinan, 250061,
    Shandong Province, China%
  }
  }
\usepackage{amsopn}


%% file: macros2e.tex
\newcommand{\realpart}{\mbox{$\mathrm{Re}$}}